\documentclass[12pt,preprint]{aastex}
\usepackage{graphics}
\shorttitle{Nuclear Reaction Rate Library Effects on Novae}
\shortauthors{Starrfield et al.}

\begin{document}
\title{The Effects of the $pep$ Nuclear Reaction and Other Improvements in the Nuclear Reaction Rate Library on
Simulations of the Classical Nova Outburst}

\author{S. Starrfield\altaffilmark{1}, C. Iliadis\altaffilmark{2},
W. R. Hix\altaffilmark{3}, F. X. Timmes\altaffilmark{1}, W. M.
Sparks\altaffilmark{4}}

\altaffiltext{1}{School of Earth and Space Exploration, Arizona
State University, Tempe, AZ 85287-1404:sumner.starrfield@asu.edu;
fxt44@mac.com} \altaffiltext{2}{Department of Physics and
Astronomy, University of North Carolina, Chapel Hill,
NC27599-3255:} \altaffiltext{3}{Physics Division, Oak Ridge
National Laboratory, Oak Ridge, TN 37831-6354 \& Department of
Physics and Astronomy, University of Tennessee, Knoxville, TN
37996-1200 :raph@ornl.gov} \altaffiltext{4}{Science Applications
International Corporation, San Diego CA, 92121 \& X-4, Los Alamos
National Laboratory, Los Alamos, NM, 87545:wms@lanl.gov}

\begin{abstract}
Nova explosions occur on the white dwarf (WD) component of a
Cataclysmic Variable binary stellar system which is accreting
matter lost by its companion.  When sufficient material has been
accreted by the WD, a thermonuclear runaway (TNR) occurs and
ejects material in what is observed as a Classical Nova explosion.
We have continued our studies of TNRs on 1.25M$_\odot$ and
1.35M$_\odot$ WDs (ONeMg composition) under conditions which
produce mass ejection and a rapid increase in the emitted light,
by examining the effects of changes in the nuclear reaction rates
on both the observable features and the nucleosynthesis during the
outburst. In order to improve our calculations over previous work,
we have incorporated a modern nuclear reaction network into our
one-dimensional, fully implicit, hydrodynamic computer code. We
find that the updates in the nuclear reaction rate libraries
change the amount of ejected mass, peak luminosity, and the
resulting nucleosynthesis.  Because the evolutionary sequences on
the 1.35M$_\odot$ WD reach higher temperatures, the effects of
library changes are more important for this mass.  In addition, as
a result of our improvements, we discovered that the $pep$
reaction ($p + e^{-} +p \rightarrow d + \nu$) was not included in
our previous studies of CN explosions (or to the best of our
knowledge those of other investigators). Although the energy
production from this reaction is not important in the Sun, the
densities in WD envelopes can exceed $10^4$ gm cm$^{-3}$ and the
presence of this reaction increases the energy generation during
the time that the $p-p$ chain is operating. Since it is only the
$p-p$ chain that is operating during most of the accretion phase
prior to the final rise to the TNR, the effect of the increased
energy generation is to reduce the evolution time to the peak of
the TNR and, thereby, the accreted mass as compared to the
evolutionary sequences done without this reaction included.  As
expected from our previous work, the reduction in accreted mass
has important consequences on the characteristics of the resulting
TNR and is discussed in this paper.
\end{abstract}


\section{Introduction}

The observable consequences of accretion onto white dwarfs (WDs)
in Close Binary stellar systems include the Classical (CN),
Symbiotic, and Recurrent Nova (RN) outbursts, and the possible
evolution of the Super Soft, Close Binary, X-ray Sources (SSS) to
Type Ia Supernovae (SNe Ia) explosions(Starrfield et al. 2004).
This diversity of phenomena occurs because of differences in the
properties of the secondary star, the mass of the WD, the stage of
evolution of the binary system (the luminosity of the WD and the
rate of mass accretion onto the WD), and the binary
characteristics (orbital separation and mass ratio).

A CN explosion occurs in the accreted hydrogen-rich envelope on
the {\it low-luminosity} WD component of a Cataclysmic Variable
(CV) system. Gas is lost by the secondary star and accreted by the
WD.  One dimensional (1D) hydrodynamic studies, which follow the
evolution of the material falling onto the WD from a bare core to
the explosion, show that the envelope grows in mass until it
reaches a temperature and density at its base that is sufficiently
high for ignition of the hydrogen-rich fuel to occur. Both
observations of the chemical abundances in CN ejecta and
theoretical studies of the consequences of the thermonuclear
runaway (TNR) in the WD envelope strongly imply that mixing of the
accreted matter with core matter occurs at some time during the
evolution to the peak of the explosion. How and when the mixing
occurs is not yet known (for discussions, see Gehrz et al. 1998:
G1998; Starrfield 2001; Starrfield, Iliadis, and Hix 2008:S2008).

If the bottom of the accreted layer is sufficiently degenerate and
well mixed with core material, then a TNR occurs and explosively
ejects core plus accreted material in a {\it fast} CN outburst.
The evolution of nuclear burning on the WD, and the total amount
of mass that it accretes and ejects depends upon: the mass and
luminosity of the underlying WD, the rate of mass accretion onto
the WD, the chemical composition in the reacting layers (which
includes the metallicity of the CV system), the convective history
of the envelope, and the outburst history of the system.

The {\it observed} levels of enrichment of elements ranging from
carbon to sulfur in CNe ejecta confirm that there is significant
dredge-up of matter from the core of the underlying WD which
enable CNe to contribute to the chemical enrichment of the ISM.
Moreover, extensive studies of CNe with IUE and the resulting
abundance determinations reveal the existence of both
oxygen-neon-magnesium (ONeMg) WDs and carbon-oxygen (CO) WDs in CN
systems (G1998). Therefore, CNe participate in the cycle of
Galactic chemical evolution in which dust grains and metal
enriched gas in their ejecta, supplementing those of supernovae,
AGB stars, and WR stars, are a source of the odd numbered, light
and intermediate mass, isotopes (and possibly other elements) in
the Interstellar Medium (ISM). Once in the diffuse gas, this
material is eventually incorporated into young stars and planetary
systems during star formation. CNe are predicted to be the major
source of $^{15}$N and $^{17}$O in the Galaxy and may contribute
to the abundances of other isotopes such as $^7$Li, $^{26}$Al and
$^{31}$P (Jos\'e and Hernanz 1998; G1998). Theoretical studies
predict that the mean mass returned by a CN outburst to the ISM is
$\sim 2\times 10^{-4}$ M$_\odot$ (G1998). Using the
observationally inferred CN rate of 35$\pm$11 per year in our
Galaxy (Shafter 1997), it follows that CNe introduce $\sim 7\times
10^{-3}$ M$_\odot$ yr$^{-1}$ of processed matter into the ISM.  It
is likely, however, that this value is a lower limit (G1998).
Recent reviews can be found in G1998, Starrfield (2001), and
S2008.

Infrared (IR) observations of the epoch of dust grain formation in
the expanding shells of CNe have confirmed that some CNe form
amorphous carbon grains, SiC grains, hydrocarbons, and oxygen-rich
silicate grains in their ejecta (some CNe form all these in the
same outburst), suggesting that a fraction of the pre-solar grains
recently identified in meteoritic material (Zinner 1998) may come
from CNe (G1998; Amari et al. 2001; Jos\'e et al. 2004; S2008; but
see also Nittler \& Hoppe 2005).

Finally, and most important to the studies in this paper, the
predictions of the 1D hydrodynamic CN simulations are directly
affected by the nuclear reactions that both determine the
production of the various isotopes and also produce the energy
that drives the ejection of the material and the shape of the
light curve. In addition, the temperatures reached around the peak
of the TNR sample the regimes of nuclear experiments where the
cross sections can be measured directly in the laboratory.
Moreover, the rates in the libraries can be tested under the same
conditions in which they were measured in the laboratory; no
extrapolations are necessary.  Therefore, over the years we have
used a variety of nuclear reaction rate libraries and determined
their influence on the properties of the outburst and the
resulting nucleosynthesis. In separate papers we have studied the
influence of various nuclear reactions on a subset of the
properties of the outburst by post-processing the results of
hydrodynamic studies (Parete-Koon, et al.~2003; Hix, et al.~2000,
2003; Iliadis, et al.~2002).  Here, we continue this work by
computing a new series of evolutionary sequences with a recent
nuclear reaction rate library.

In the next section we briefly describe both the changes to the
NOVA code and the four reaction rate libraries that are used for
the calculations reported in this paper. In the following section,
we report on the results of our new calculations.  We continue, in
Section 4, with a discussion of the resulting nucleosynthesis, and
end with a summary and discussion.

\section{The Hydrodynamic Computer Code and Nuclear Reaction Rate
Libraries}

Over the past few years we have been improving the physics in NOVA
and then determining the effects of the improved physics on
simulations of the CN outburst (S1998; Starrfield et al. 2000,
S2000, S2008).  NOVA is a 1D, Lagrangian, fully implicit,
hydrodynamic computer code that incorporates a large nuclear
reaction rate network.  It is described in detail in S1998; S2000;
and references therein. As reported in those papers, we have found
that improving the opacities, equations-of-state, and the nuclear
reaction rate library have had important effects on both the
energetics and the nucleosynthesis. Similar results have been
reported by the Barcelona group (Hernanz and Jos\`e 2000, and
references therein).  We have continued to explore the effects of
improving the reaction rates used in the calculations on the
evolution of the CN outburst.  In this paper we compare our
earlier studies to new simulations using a reaction rate library
of Iliadis which is current as of August 2005 (hereafter I2005).
In addition, NOVA is continuously being updated and for the work
reported in this paper we have made one major change and numerous
minor changes.

The major change is that we no longer use the nuclear reaction
network of Weiss and Truran (1990: WT1990) but have switched to
the modern nuclear reaction network of Hix and Thielemann (1999:
HT1999; see also Parete-Koon et al. 2003).  While both networks
utilize reaction rates in the common REACLIB format and perform
their temporal integration using the Backward Euler method
introduced by Arnett and Truran (1969), two important differences
are evident. First, WT1990 implemented a single iteration,
semi-implicit backward Euler scheme, which had the advantage of a
relatively small and predictable number of matrix solutions, but
allowed only heuristic checks that the chosen timestep resulted in
a stable or accurate solution.  In contrast, HT1999 implemented an
iterative, fully implicit scheme, repeating the Backward Euler
step until convergence is achieved.  The iterations provide a
measure both of the stability and the accuracy of the solution.
Moreover, if convergence does not occur within a reasonable number
of iterations, then the timestep is subdivided into smaller
intervals until a converged solution can be achieved. Therefore,
the fully implicit backward Euler integration can respond to
instability or inaccuracy in a way that is impossible with the
semi-implicit backward Euler approach. As a result, the fully
iterative approach can often safely employ larger time steps than
the semi-implicit approach, obviating the speed advantage of the
semi-implicit method's smaller number of matrix solutions per
integration step. In addition to the changes in the nuclear
reaction network and library, we now use the weak and intermediate
screening equations from Graboske et al. (1973) instead of the
framework of Salpeter (1954) as described in Cox and Giuli (1968).

Finally, the HT1999 network employs automated linking of reactions
in the data set to the species being evolved. This is in contrast
to the manual linking employed by WT1990 and many older reaction
networks.  The automated linking helps to avoid implementation
mistakes, as we discovered while performing tests of NOVA in order
to understand the source of differences in the results of the
simulations between the two versions of the code which used the
same reaction rate library but different nuclear reaction networks
(plus other differences).  We found that while the REACLIB dataset
used in prior studies (Politano et al.\ 1995: P1995; S1998;
S2000), included the $pep$ reaction ($p + e^{-} +p \rightarrow d +
\nu$: Schatzman 1958; Bahcall \& May 1969), it was not linked to
abundance changes (or the resulting energy generation) in the
WT1990 network. While for Solar modeling energy generation from
the {\it pep} reaction is unimportant (but not the neutrino
losses: Rolfs and Rodney 1988), in the WD envelope the density can
reach, or exceed, values of $10^4$ gm cm$^{-3}$ which, in turn,
increases the rate of energy generation over the simulations done
without the {\it pep} reaction included (Starrfield et al.~2007).
The increased energy generation reduces the amount of accreted
material since the temperature rises faster per gram of accreted
material. The effect of changes in the rate of energy generation
on simulations of the CN outburst is discussed in detail in S1998.
Given a smaller amount of accreted material at the time when the
steep temperature rise begins in the TNR, the nuclear burning
region is less degenerate and, therefore, the peak temperatures
are lower compared to models evolved with the same nuclear
reaction rate library used in our previous studies (see the
evolution sections below). To the best of our knowledge, none of
the previous studies of TNR's in WD envelopes have included this
reaction.

Another difference, between this work and our previous studies, is
that we do not initiate nuclear burning until the temperatures
have reached 9 million degrees in a given mass zone. In our
earlier studies, done with the WT network, nuclear burning was
initiated at 4 million degrees.  We made this change because the
reaction rates in the latest libraries are not fit to temperatures
below about 10 million degrees and, for temperatures of about 8
million degrees (and lower) some of the rates begin increasing
rapidly and unrealistically. Our test runs found that 9 million
degrees was a good cut-off value. Fortunately, this has almost no
effect on the evolution. We find, for example, that for the
sequence done with the  Iliadis 2005 reaction rate library at
1.35M$_\odot$, nuclear burning does not start until the sequence
has evolved for $7.1 \times 10^3$ yr (and the accreted material at
that time, $\sim 10^{-6}$M$_\odot$, has reached down from the
surface [mass zone 95] to mass zone 81 [$1.1 \times
10^{-6}$M$_\odot$]) compared to the total accretion time of $1.8
\times 10^5$ yr (where the accreted material [see Table 2] has
reached down to mass zone 63 [$ 2.1 \times 10^{-5}$M$_\odot$]).
Therefore, there is no nuclear burning in the accreted material
for only $\sim$4\% of the evolution time and, given that the $p-p$
chain is operating at this time, only a small fraction of the
total $\it nuclear$ energy production is neglected.

However, this also means that the outermost mass zones, which have
temperatures below 9 million degrees until near the peak of the
outburst (when the energy and products of nuclear burning are
brought to the surface by convection), do not experience nuclear
burning during the accretion phase of the outburst. Therefore,
when these layers are mixed into the nuclear burning layers near
the peak of the outburst they inject a larger amount of
unprocessed nuclei into the TNR than found in our earlier
simulations. In order to understand the effects of this
difference, we redid the earlier calculations with the older
reaction rate libraries (see below).

In this paper we evolve seven different sequences using the same
initial conditions but four different reaction rate libraries for
each of the two WD masses.  We report the results in the tables
described in the next section.  The first library we use includes
the rates from Caughlan and Fowler (1988) and Thielemann et al.
(1987, 1988). They were compiled by Thielemann, made available to
Truran and Starrfield, and used for the calculations reported in
WT1990 and those in Politano et al. (1995: P1995). The first
sequence (labeled P1995A), uses the P1995 library, the WT1990
network, and none of the updates listed in the last section. The
second sequence (labeled P1995B) uses the latest version of the
NOVA code, the HT1999 network, the P1995 library, but the $pep$
reaction is not included.  A comparison of the results from these
two sequences shows the results of updating the code. Most of the
differences can be attributed to our use of the Graboske et al.
(1973) screening in the latest version.

The third sequence (labeled P1995C) is identical to sequence 2
except that it includes the $pep$ reaction and shows how including
the $pep$ reaction changes the results of the evolution.  The next
three sequences are done with three different reaction rate
libraries. The second library (labeled S1998) uses an updated
reaction rate library which contains new rates calculated,
measured, and/or compiled by Thielemann and Wiescher. A discussion
of the improvements is provided in S1998. The third library
(labeled I2001) is described in Iliadis et al. (2001) and was used
for the simulations reported in Starrfield et al. (2001).  The
fourth library (labeled I2005A) is the August 2005 library of
Iliadis and the results of calculations done with this library are
given in this paper.  We label it ``This Work'' in the plots.
These three sequences include the $pep$ reaction.  Finally, there
is one last sequence (labeled I2005B) which is identical to I2005A
except that the $pep$ reaction is {\it not} included.  Therefore,
we can compare sequences P1995B and P1995C and I2005A and I2005B
to determine just the effects of including the $pep$ reaction,
while sequences P1995C, S1998, I2001, and I2005A show the effects
of the different reaction rate libraries on the evolution. In
order to prevent confusion, the particular library used for each
sequence, the reaction network, and whether or not the $pep$
reaction is included are listed in the comments to Tables 2.

References to many of the updated reaction rates used in
calculating sequences I2005A and I2005B are given in Table 1 along
with some comments on those rates. As required, the ground and
isomeric states of $^{26}$Al are treated as separate nuclei (Ward
\& Fowler 1980) and the communication between those states through
thermal excitations involving higher lying excited $^{26}$Al
levels is taken into account.  The required $\gamma$-ray
transition probabilities are adopted from Runkle et al. (2001).

\clearpage

\begin{deluxetable}{lll}
\tiny \tablecaption{Sources of reaction rates\tablenotemark{a}}
\tablehead{\colhead{Reaction} & \colhead{Source} &
\colhead{Comment} } \startdata $^{8}$B(p,$\gamma$)$^{9}$C &
Beaumel et al.\ 2001      & in close agreement with Trache et al.\
2002 \\ $^{11}$C(p,$\gamma$)$^{12}$N        & Tang et al.\ 2003 &
rate of Liu et al.\ 2003 is higher by a factor of 2\\
$^{13}$N(p,$\gamma$)$^{14}$O        & Tang et al.\ 2004 & \\
$^{14}$N(p,$\gamma$)$^{15}$O        & Champagne 2004 & based on
Runkle et al.\ 2005 \\
$^{15}$O($\alpha$,$\gamma$)$^{19}$Ne & Davids 2004          & \\
$^{17}$O(p,$\gamma$)$^{18}$F        & Fox et al.\ 2005       & \\
$^{17}$O(p,$\alpha$)$^{14}$N & Chafa et al.\ 2005 & \\
$^{17}$F(p,$\gamma$)$^{18}$Ne       & Iliadis et al.\ 2008 & with
information from Bardayan et al.\ 2000 \\
$^{18}$F(p,$\gamma$)$^{19}$Ne           & de S\'{e}r\'{e}ville et
al.\ 2005     & \nl $^{18}$F(p,$\alpha$)$^{15}$O & de
S\'{e}r\'{e}ville et al.\ 2005        & \\
$^{18}$Ne($\alpha$,p)$^{21}$Na          & Chen et al.\ 2001 & \\
$^{19}$Ne(p,$\gamma$)$^{20}$Na      & Vancraeynest et al.\ 1998 &
\\ $^{23}$Na(p,$\gamma$)$^{24}$Mg      & Rowland et al.\ 2004 & \\
$^{23}$Na(p,$\alpha$)$^{20}$Ne      &Rowland et al.\ 2004 & \\
$^{25}$Al(p,$\gamma$)$^{26}$Si          & Iliadis et al.\ 2008 &
based on Parpottas et al.\ 2004 and Bardayan et al.\ 2002 \\
$^{29}$Si(p,$\gamma$)$^{30}$P & Iliadis et al.\ 2008 & \\
$^{30}$Si(p,$\gamma$)$^{31}$P       & Iliadis et al.\ 2008 & \\
\enddata
\tablenotetext{a}{Rates of most other reactions not listed above
are adopted from Caughlan \& Fowler 1988, Angulo et al.\ 1999, and
Iliadis et al.\ 2001.}
\end{deluxetable}

\clearpage

Other changes to NOVA include the use of the analytic fitting
formulas of Itoh et al. (1996) for the neutrino energy loss rates
from pair ($e^+ + e^- \rightarrow \nu_e + {\bar \nu}_e$), photo
($e^{\pm} + \gamma \rightarrow e^{\pm} + \nu + \bar{ \nu}_e$),
plasma ($\gamma_{{\rm plasmon}} \rightarrow \nu_e + \bar{
\nu}_e$), bremsstrahlung ($e^- + A^Z \rightarrow  e^- + A^Z +
\nu_e + \bar{\nu} _e$), and recombination ($e^-_{\rm continuum}
\rightarrow e^-_{\rm bound} + \nu_e + \bar{ \nu}_e$) processes. As
stellar evolution codes generally require derivative information
for the Jacobian matrix, our implementation of the Itoh et al.
(1996) fitting formulas (available online from cococubed.asu.edu)
returns the neutrino loss rate and its first derivatives with
respect to temperature, density, $\bar{A}$ (average atomic weight)
and $\bar{Z}$ (average charge). Finally, we assume a value of  2
for the mixing-length to scale height ratio ($l/H_p$). There are
additional and numerous small changes to NOVA that had minimal
effects on the simulations to be described in the next section.

\section{The Initial Conditions and Evolutionary Results}

Our initial models are complete 1.25M$_\odot$ and 1.35M$_\odot$
WDs discretized into 95 zones. This is the same number used in our
previous studies (P1995, S1998, and S2000). We assume that the
material being accreted from the donor star is of Solar (Anders
and Grevesse 1989) composition and that it has already mixed with
the core material so that the actual accreting composition in this
study is 50\% Solar and 50\% ONeMg material (We use the ONeMg
composition of Arnett and Truran 1969.).  The use of this
composition affects the total amount of accreted mass at the peak
of the TNR since it has a higher opacity than if no mixing were
assumed (S1998; Jos\'e et al. 2007). The initial (Solar and ONeMg
mixed) abundances by mass are given in column 7 of Table 4.

We use an initial WD luminosity of either $\sim 3 \times 10^{-3}$
L$_\odot$ (1.25M$_\odot$) or  $\sim 4 \times 10^{-3}$ L$_\odot$
(1.35M$_\odot$).  We use a smaller value for the accretion rate
(than in S1998), $1.6 \times 10^{-10}$M$_\odot$yr$^{-1}$, in order
to accrete the largest amount of mass possible for a given WD
mass. This mass accretion rate is 5 times lower than the lowest
rate used in either S1998 or S2000 and was chosen to maximize the
amount of accreted material given the increased energy generation
from including the $pep$ reaction.  Studies of accretion onto WDs
demonstrate that the results of the evolution depend strongly on
the initial WD luminosity and mass accretion rate (c.f., Yaron et
al. 2005; G1998; S1998; S2000; S2008, and references therein).

The results of our evolutionary calculations are given in Tables 2
through 5.  Tables 2 and 3 give the initial conditions and
evolutionary results for both WD masses while Tables 4 and 5 give
the abundances of the ejected material (by mass) for the 8
different simulations done with the $pep$ reaction included.  The
numerical factor that multiplies the abundances is given in the
first column next to the isotope designation. The rows in Tables 2
and 3 are the reaction rate library, the accretion time to the TNR
($\tau$$_{\rm acc}$), the accreted mass (M$_{\rm acc}$), peak
temperature in the TNR (T$_{peak}$), peak rate of energy
generation during the TNR ($\epsilon_{\rm nuc-peak}$), peak
luminosity (L$_{\rm peak}$), peak effective temperature (T$_{\rm
eff-peak}$), ejected mass (M$_{\rm ej}$), and the peak expansion
velocity after the radii of the surface layers have reached $\sim
10^{13}$cm (V$_{\rm max}$). By this time the outer layers are
optically thin, have far exceeded the escape velocity at this
radius, and there is no doubt that they are escaping.

As mentioned in the last section, the first three sequences
(P1995A, P1995B, and P1995C) employ the P1995 reaction rate
library. As shown in Tables 2 and 3, the three different sequences
using the P1995 library provide a clear picture both of the impact
of the $pep$ reaction as well as the other updates to the NOVA
code. Since neither sequence P1995A nor P1995B included the $pep$
reaction, the differences between these two sequences shows only
the impact of the updates to the NOVA code, which are noticeable
but generally small.  Most of these differences can be attributed
to our use of the Graboske et al. (1973) weak and intermediate
screening in this paper and not in earlier papers.

In contrast, much larger differences are seen when sequence P1995B
is compared to P1995C which does include the $pep$ reaction but is
otherwise identical.  This conclusion is reinforced by comparing
sequences I2005A and I2005B (both calculated with the latest
library [I2005] and current version of NOVA) since I2005A includes
the $pep$ reaction and it is not included in I2005B.  Tables 2
(1.25M$_\odot$) and 3 (1.35M$_\odot$) show that for both WD masses
the largest change in the results of the evolution occurs with the
inclusion of the $pep$ reaction.  If we compare rows P1995B and
P1995C or rows I2005A or I2005B, then the increase in energy
production from adding the $pep$ reaction to the network results
in a significant decrease in both accretion time and accreted
mass. Because there is less accreted mass on the WD at the time of
the TNR, all the peak values are smaller. However, the effects are
much more important for the 1.35M$_\odot$ evolution than for the
1.25M$_\odot$ evolution. This is because the density is higher in
the more massive WD at the beginning of the TNR and, therefore,
the $pep$ reaction provides more energy ($\epsilon \sim \rho^2$).
Interestingly enough, the peak temperatures reached in the two
1.35M$_\odot$ simulations without the $pep$ reaction (P1995B and
I2005B) exceed $5 \times 10^8$K which is sufficiently high for CNO
breakout (the $^{14}$O($\alpha$,$\gamma$) and
$^{15}$O($\alpha$,$\gamma$) reactions) to occur (see below).
Unfortunately, evolution without the necessary physics of the
$pep$ reaction included is not realistic and we will have to look
elsewhere for initial conditions that produce sufficiently high
temperatures for breakout.

If we compare the results for the four sequences with the $pep$
reaction included (P1995C, S1998, I2001, I2005A), we see that
changes in the nuclear reaction rate library produce differences
in ejected mass and peak luminosity for both the 1.25M$_\odot$ and
1.35M$_\odot$ evolutionary sequences. The results of the
evolutionary sequences show that because the WD mass is larger and
the radius is smaller for 1.35M$_\odot$, the mass zones where the
TNR occurs reach higher densities and higher peak temperatures
than do the sequences at lower WD mass (Starrfield 1989; S2008;
Yaron et al. 2005).  At 1.35M$_\odot$ the sequence done with the
latest reaction rate library (I2005A) accretes and ejects the
lowest amount of mass moving at the lowest ejection velocities. In
addition, the peak luminosity and effective temperature is lowest
for the calculation done with this library. The amount of ejected
mass and the ejection velocities are in disagreement with the
observations (S2008).

Table 2 shows that only about 25\% of the accreted material is
ejected in the explosive phase of the outburst at 1.25M$_\odot$
and Table 3 shows about 60\% of the accreted material is ejected
at 1.35M$_\odot$.  This is a common feature of our 1D hydrodynamic
simulations (G1998; S2008). The material that is not ejected
returns to quasistatic equilibrium on the WD and stays luminous
and hot with radii exceeding $10^9$cm.  X-ray studies of this
phase of evolution for a CN in outburst indicate that we are
observing a hot, luminous stellar atmosphere (Petz et al. 2005;
Ness et al. 2007) just as in the Super Soft X-ray Binaries such as
CAL 83 (Kahabka \& van den Heuvel 1997; Lanz et al. 2005). The
predicted time required to burn the remaining envelope material
and return the CN to quiescence can exceed 100 yr (Starrfield
1989) which is not observed (Orio 2004). It has been proposed that
the remaining material is ejected via radiation pressure driven
mass loss on short timescales (Starrfield 1979; MacDonald, Truran,
and Fujimoto 1985; Starrfield et al. 1991). Nevertheless, some of
the accreted envelope may actually be burnt to helium enriched
material and become part of the material ejected in the next CN
outburst (Krautter et al. 1996). However, the amount of accreted
material that is not ejected suggests that it is insufficient to
counteract the amount of WD core mass lost in the outburst. As a
result, the WD is losing mass as a result of the CN outburst and
CNe cannot be the progenitors of Supernovae of Type Ia.

Figure 1 shows the variation of temperature with time for the zone
where peak conditions in the TNR occur in the 1.25M$_\odot$
evolutionary sequences. In this figure and all other figures we
plot only the four simulations done with the $pep$ reaction
included.  The specific evolutionary sequence is identified on the
plot and in the caption. The reference to the nuclear reaction
rate library used for that calculation is given in the caption for
Figure 1 and indicated on the figure. The designation ``This
Work'' refers to the I2005A sequence as described in the tables
and discussed earlier. The time coordinate is chosen to clearly
show the rise and decline time of each evolutionary sequence.
Interestingly enough, the rise time and peak temperature are
nearly the same for all four sequences at 1.25M$_\odot$.

Figure 2 shows the same plot for the sequences at 1.35M$_\odot$.
Here we see differences between the four simulations.  Peak
temperature drops from about 413 million degrees to 392 million
degrees and peak nuclear energy generation drops by about a factor
of 2 from the oldest library to the newest library ($8.4 \times
10^{17}$erg gm$^{-1}$s$^{-1}$ to $4.4 \times 10^{17}$erg
gm$^{-1}$s$^{-1}$). The temperature declines more rapidly for the
sequence (P1995C) computed with the oldest reaction library
(P1995) because there was a larger release of nuclear energy
throughout the evolution so that the overlying zones expanded more
rapidly and the nuclear burning region cooled more rapidly than in
the other sequences. In contrast, the newest library, showing the
smallest expansion velocities, cools slowly. There is a factor of
two difference in the time coordinate used for Figures 1 and 2
because the simulations at 1.35M$_\odot$ evolve much more rapidly
near the peak of the TNR than those at 1.25M$_\odot$. This is a
direct result of the higher gravity and higher degeneracy in the
nuclear burning region of the more massive WDs.

The evolution of the total nuclear energy generation in the
nuclear burning layers (in Solar units: L/L$_\odot$) as a function
of time for each mass is shown in Figures 3 (1.25M$_\odot$) and 4
(1.35M$_\odot$). The time coordinates are the same as those used
in Figures 1 and 2, respectively. Again, there is hardly any
difference at the lower WD mass but at 1.35M$_\odot$ the peak for
the calculation done with the latest library is definitely lower
than seen in the earlier libraries.

Figures 5 and 6 show the variation of the effective temperature
(T$_{\rm{eff}}$) with time as the layers begin their expansion. We
plot the results with the same time coordinates as in Figures 1
and 2 and these plots show how rapidly the energy and
$\beta^+$-unstable nuclei reach and heat the surface layers. Note
that peak T$_{\rm{eff}}$ occurs when the WD radius is still small
and earlier in the evolution than when peak luminosity occurs.
The large amplitude oscillations seen in the sequences using the
older libraries, and not in that from the latest library, are
caused by the intense and rapid heating of the surface layers.
They expand, cool, collapse back onto the surface, and expand
again. The outer layers are still deep within the gravitational
potential well of the WD (since hardly any expansion has occurred
at the time of the oscillations) and so the ``quasi''-period is
that of the free-fall time for the underlying WD.  After a few
seconds the outer layers are expanding sufficiently rapidly, have
cooled, and the oscillations cease. They are not present in the
sequence using the latest library because surface heating is less
important.  The outburst evolves more gradually and the star has
started to expand by the time that the $\beta^+$-unstable nuclei
reach the surface. This can also be seen in Figure 7
(1.35M$_\odot$) which shows the velocity of the surface layers as
a function of time around the time of peak temperature in the
nuclear burning region.

Figures 8 and 9 show the variation with time of the surface
luminosity (for the first 11 hours of the TNR) for each of the two
WD masses.  The intense heat from the $\beta^+$-unstable nuclei
causes the luminosity to become super-Eddington and the layers
begin expanding. However, they are still deep within the potential
well of the WD and oscillate for a few seconds.  In contrast, the
sequence done with the latest library does not become as luminous
and the initial oscillations exhibit a much smaller amplitude.
These figures imply that if we could observe a CN sufficiently
early in the outburst, then it would be super-Eddington and
emitting soft X-rays.  However, by the time a CN is typically
discovered its luminosity has declined to below Eddington. In
contrast to this result, however, the IUE observations of LMC 1991
showed that it was super-Eddington for more than 10 days (Schwarz
et al. 2001) a result that is not predicted by any existing CN
simulations (S2008).

The initial spike (at a time of about 100 s) is caused by a
slowing of the expansion as the energy produced by the
$\beta^+$-decays decreases. After this time, expansion and cooling
of the outer layers causes the opacity to increase and radiation
pressure then accelerates the layers outward.  The continuous flow
of heat from the interior, combined with the increase in opacity,
causes another increase in luminosity until the peak is reached.

\section{Nucleosynthesis}

In this section we present the predicted ejecta abundances for the
four sequences at each WD mass done with the $pep$ reaction
included. Because it is a necessary piece of the $p-p$ reaction
chain, calculations done without it included in the reaction
network are not realistic and we do not report the abundance
results for the three sequences done without the $pep$ reaction at
each WD mass. However, we do provide two plots which show the
effects on the abundances of not including the $pep$ reaction and
discuss them below.  The results for each nucleus in our nuclear
reaction network are given as mass fraction in Table 4
(1.25M$_\odot$) and Table 5 (1.35M$_\odot$). The factor
multiplying each isotope can be found just to the right of the
isotopic designation.  Note that the right hand column in Table 4
is the initial abundance of the given nucleus (we do not repeat
this column in Table 5).

In order to more clearly show which nuclei are produced by CNe
explosions, in Figures 10 and 11 we plot the stable, ejected
nuclei divided by the Anders and Grevesse (1989) Solar abundances.
In both figures the $x$-axis is the atomic mass number. The
$y$-axis is the logarithmic ratio of the ejecta abundance divided
by the Solar abundance of the same nucleus.  The most abundant
isotope of a given element is marked by an asterisk and isotopes
of the same element are connected by solid lines and labelled by
the given element. These plots are patterned after similar plots
in Timmes et al. (1995).  They show for both WD masses that we
predict that $^{15}$N, $^{17}$O, and $^{31}$P are overproduced by
a factor of $10^4$ in CNe ejecta. There are other nuclei that are
overproduced by factors of a thousand and could be important for
CN nucleosynthesis.  In Figures 12 (1.25M$_\odot$) and 13
(1.35M$_\odot$) we show the ratio of the ejected abundances for
the simulation (I2005A) done with the $pep$ reaction (using the
I2005 reaction rate library) compared to a simulation (I2005B)
done without the $pep$ reaction (also using the I2005 library).
The plot style is the same as for Figures 10 and 11 as described
above except that the $y$-axis is {\it linear} and not
logarithmic.

The initial abundance of $^1$H is 0.365. (This value is half the
solar abundance of Anders and Grevesse [1989].) The hydrogen
abundance in the ejected gases for the 1.25M$_\odot$ I2005A
sequence has declined to $\sim$0.31. This decline of $\sim$0.05 in
mass fraction results in a total energy production from proton
captures of $\sim 4 \times 10^{46}$ erg which agrees with the
values typically quoted for observed CN explosions (Starrfield
1989; G1998; S2008). Interestingly, the ejecta abundance of $^4$He
decreases slightly as the reaction rate library is improved and
the smallest increase occurs in the calculations done with the two
most recent libraries. A $^4$He ejecta abundance of 0.16 is far
smaller than the values typically quoted for observed CN ejecta
(G1998 and references therein).  We, therefore, support the
speculation of Krautter et al. (1996; see also S1998 and S2000)
that the large amount of helium observed in CN ejecta implies: (1)
that most of the ejected helium was mixed up from the outer layers
of the WD by the TNR; and (2) that it was actually produced in
previous CN outbursts and subsequent nuclear burning on the WD.

Turning to the more massive nuclei, the abundances of $^{12}$C and
$^{13}$C drop by about a factor of two from the oldest to the
newest reaction rate libraries while $^{14}$N increases by about a
factor of two and $^{15}$N declines by slightly less than a factor
of two. Note that the abundance of $^{15}$N far exceeds that of
$^{14}$N and it is likely that the nitrogen observed in CN ejecta
is mostly $^{15}$N rather than $^{14}$N. Therefore, our
speculation about helium may also hold true for nitrogen.  The
observed nitrogen is probably $^{15}$N produced in previous
outbursts, mixed into the newly accreted material, and then
ejected during the current CN outburst.

  Similarly, $^{16}$O declines by about 30\% while $^{17}$O
increases by more than a factor of two.  In fact, $^{17}$O is the
most abundant of the CNO nuclei in the ejecta and the abundances
of $^{15}$N and $^{17}$O exceed those of the even-numbered
isotopes. Finally, the C/O ratio in the ejecta drops from about
30\% to about 8\% from the earliest to the latest library.  Other
interesting nuclei at this WD mass are $^{18}$O which drops about
a factor of 10 as the library is improved, $^{22}$Na whose
abundance remains virtually unchanged as the library is improved,
$^{24}$Mg which drops a factor of 3 and is severely depleted from
its initial abundance, and both $^{26}$Al and $^{27}$Al which drop
by about a factor of two in abundance.

We also find that most of the higher mass nuclei ($^{40}$Ca is the
most massive nucleus in our network), are all produced by the TNR
in the outburst (see Table 4).  Interestingly, the abundances of
$^{28}$Si, $^{29}$Si, and $^{30}$Si are largest in the
calculations done with the latest library while the abundances of
$^{33}$S,  $^{34}$S, and  $^{35}$Cl decrease in the latest
library.  Finally, $^{36}$Ar is depleted by the outburst (its
final abundance is less than the initial abundance) in all
sequences except P1995C.

The ejecta abundance results for TNRs on 1.35M$_\odot$ WDs are
given in Table 5 for the same reaction rate libraries used in the
study at lower WD mass. Hydrogen is depleted by a larger amount at
this WD mass than for the 1.25M$_\odot$ sequences resulting in a
total energy production from proton captures of $\sim 3 \times
10^{46}$ erg. This value is smaller than in the lower mass
sequence because the 1.35M$_\odot$ sequences accrete less mass. As
in the sequences at 1.25M$_\odot$, the helium abundance in the
ejecta is small compared to the observed helium abundances in CN
ejecta and the results at this WD mass also support our prediction
that the accreted material mixed with the outer layers of the WD
at some time during the outburst.  We emphasize, in addition, that
the large helium abundances observed in recurrent novae such as U
Sco or V394 CrA imply that mixing with the WD has occurred in
these systems even if the total CNO abundances in their ejecta are
not dramatically enriched over solar (Shore et al. 1991).

Examining the behavior of the individual abundances, we see that
$^{12}$C and $^{13}$C are virtually unchanged by the updated
reaction rates. In contrast, the abundance of $^{14}$N nearly
doubles and that of $^{15}$N decreases by a factor of two going
from the earliest to the latest reaction rate library.  $^{16}$O
doubles in abundance while $^{17}$O grows by a factor of 60 and
becomes the most abundant of the CNO nuclei in the ejecta. For
this WD mass and the latest library, the C/O ratio is 0.12.  The
abundance of $^{18}$O declines by nearly a factor of 5 and the
abundances of $^{18}$F and $^{19}$F also decline by large factors
in the sequence done with the latest library.

The initial abundance of $^{20}$Ne in all four sequences is 0.25
(see Table 4) so that it is depleted by a smaller amount in the
calculations done with the latest library.  The abundance of
$^{22}$Na decreases with the library update and $^{24}$Mg is
severely depleted by the TNR. In fact, all the Mg isotopes are
depleted in the calculations done with the latest library.  In
contrast, the ejecta abundance of $^{26}$Al is unchanged by the
changes in the reaction rates while the abundance of $^{27}$Al
drops by a factor of two. We also find, contrary to a conclusion
in Politano et al. (1995), that the amount of $^{26}$Al ejected is
virtually independent of WD mass.

All the Si isotopes ($^{28}$Si, $^{29}$Si, and $^{30}$Si) are
enriched in the calculations done with the latest library and
$^{29}$Si, and $^{30}$Si are more abundant in the 1.35M$_\odot$
simulations than the 1.25M$_\odot$ simulations.  Other nuclei
whose abundances are largest in the calculations done with the
latest library are $^{31}$P and $^{32}$S.  These nuclei are also
more abundant at the higher WD mass.  In addition, while the
ejecta abundance of $^{33}$S does not depend on the reaction rate
library, it is nearly 30 times more abundant in the calculations
done with the more massive WD.  Finally, we note that while the
ejecta abundances of $^{34}$S, $^{35}$Cl, $^{36}$Ar, and $^{40}$Ca
have all declined as the reaction rate library has been improved,
we predict that they will be produced in a nova TNR since their
final abundances exceed the initial abundances.

The effects of including the $pep$ reaction on ejecta abundances
are shown in Figures 12 and 13 in which we plot the ratio of the
ejecta abundance of the sequence with the $pep$ reaction included
divided by the abundance from the equivalent sequence with the
$pep$ reaction not included.  Figure 12 shows that most nuclei
have a higher abundance in the 1.25M$_\odot$ sequence with no
$pep$. This is also true for the 1.35M$_\odot$ WD sequence with
the notable exception of the carbon isotopes, $^{14}$N, $^{20}$Ne,
$^{21}$Ne, and $^{32}$S. These results are as expected since the
sequence at 1.35M$_\odot$ reaches to higher temperatures.

Finally, given the high temperatures attained in the 1.35M$_\odot$
sequence without the $pep$ reaction included, we checked to
determine if breakout had occurred. We found that the total CNO
abundances decreased from their initial values in both sequences
(using the latest reaction library only). As expected, the
sequence done without the $pep$ reaction showed the most depletion
but for neither mass was it sufficiently large to be observable.

\section{Summary and Discussion}

In this paper we examined the consequences of improving the
nuclear reaction rate library on our simulations of TNRs on
1.25M$_\odot$ and  1.35M$_\odot$ ONeMg WDs.  We found that the
changes in the rates affected predictions of both the
nucleosynthesis and the observable features of the evolution such
as peak luminosity, peak effective temperature, ejected mass, and
ejecta velocities. A major change to our previous calculations,
that effects virtually all features of the predicted outburst, has
been the inclusion of the $pep$ reaction in the $p-p$ chain.  This
reaction is important during the accretion phase of the evolution
because the density of the accreting material quickly reaches
values of $\sim 10^4$ gm cm$^{-3}$.  This high a density increases
the nuclear energy generation over studies done with the $pep$
reaction absent. The increased energy generation reduces the time
to reach the TNR and, thereby, the amount of accreted material and
as a result the peak values of temperature and energy generation
are smaller than we have found in our previous studies.

If we examine the abundance predictions for the four 1.25M$_\odot$
sequences done with the $pep$ reaction included, we see that the
differences caused by improving the reaction rate library are that
the abundance of $^{12}$C declines by about a factor of two (all
abundances are given in mass fraction), $^{14}$N increases by
almost a factor of two, and $^{16}$O declines by about a factor of
1.5. Both $^{12}$C and $^{13}$C are depleted in the latest
sequence (compared to the P1995 library) as is $^{15}$N while
$^{17}$O is enriched in the calculation done with the latest
reaction rate library.  In addition, in all four sequences the
ejected oxygen exceeds carbon as found in our earlier studies.
This result continues to be puzzling in light of the production of
carbon rich dust grains in CN ejecta (G1998).  It is possible that
the carbon dust grain forming CNe occur on lower mass CO WDs which
never develop sufficiently hot nuclear burning temperatures to
deplete the carbon as compared to oxygen. As we examine the more
massive nuclei at 1.25M$_\odot$, we see that $^{26}$Al, and
$^{27}$Al are depleted in the simulations done with the latest
library while $^{32}$S is enhanced.  Interestingly, the abundance
of $^{22}$Na increased with the sequences done with the libraries
intermediate in time but then decreased to a value nearly equal to
that in the earliest library.

The effects of changing the nuclear reaction library are also
apparent for the sequences at 1.35M$_\odot$.  Both $^{12}$C and
$^{13}$C drop in abundance while $^{14}$N, $^{16}$O and $^{17}$O
increase in abundance.  We find that while the ejecta abundance of
$^{22}$Na is lowest in simulations done with the I2005 library, it
is still a factor of about 5 more abundant at 1.35M$_\odot$ than
at 1.25M$_\odot$.  The abundance of $^{26}$Al is unchanged while
that of $^{27}$Al declined by about a factor of 2.  In addition,
the abundance of $^{26}$Al is roughly constant from one WD mass to
the other while the abundance of $^{22}$Na declined by about a
factor of 2 as the WD mass increased. This is not what we reported
in earlier studies (done without the $pep$ reaction) using older
libraries where we found that the abundance of $^{26}$Al declined
as the WD mass increased.  Finally, we note that the abundance of
$^{32}$S is largest for the latest library at 1.35M$_\odot$. In
fact, it reaches 4\% of the ejected material.

In summary, the nucleosynthesis predictions from our simulations
show significant impact from improvements in the reaction rates
over the past 15 or so years.  Observable features of the models,
such as the variation of the effective temperature and luminosity
with time, and also the mass ejected, exhibit a notable influence
from changes in these rates, because of their dependence on
heating from the decays of nucleosynthesis products that have been
mixed into the outer layers.

We thank L. Bildsten, A. Champagne, R. Gehrz, J. Krautter, H.
Schatz, D. Townsley, J. Truran, and C. E. Woodward for interesting
discussions. We are grateful to the anonymous referee whose
comments improved the presentation of this paper. SS thanks J.
Aufdenberg and ORNL for generous allotments of computer time. CI
is supported in part by the U.S. Department of Energy under
Contract No. DE-FG02-97ER41041. WRH has been partly supported by
the National Science Foundation under contracts PHY-0244783 and
AST-0653376. Oak Ridge National Laboratory is managed by
UT-Battelle, LLC, for the U.S. Department of Energy under contract
DE-AC05-00OR22725. S. Starrfield acknowledges partial support from
NSF and NASA grants to ASU.

\clearpage

\begin{deluxetable}{@{}lccccccc}
\tablecaption{Initial Parameters and Evolutionary Results for
1.25M$_\odot$ White Dwarfs \tablenotemark{a}\label{1p25evol}}
\tablewidth{0pt} \tablehead{ \colhead{Sequence:} &
\colhead{P1995A\tablenotemark{b}} &
\colhead{P1995B\tablenotemark{c}} &
\colhead{P1995C\tablenotemark{d}} &
\colhead{S1998\tablenotemark{e}} &
\colhead{I2001\tablenotemark{f}} &
\colhead{I2005A\tablenotemark{g}} &
\colhead{I2005B\tablenotemark{h}}} \startdata

$\tau$$_{\rm acc}$($10^5$ yr)&5.2&5.0&3.8&3.8&3.8&3.8&5.0 \\
M$_{\rm acc}$(10$^{-5}$M$_{\odot}$)&8.2&8.0& 6.0& 6.0 & 6.1 & 6.1 &8.0\\
T$_{\rm peak}$($10^6$K)&348& 347&321  & 321 &  320& 320&347 \\
$\epsilon_{\rm nuc-peak}$(10$^{17}$erg gm$^{-1}$s$^{-1}$)&2.8 &2.7& 2.0& 2.1  & 1.3  & 1.3 &1.8\\
L$_{\rm peak}$ (10$^5$L$_{\odot}$)&4.2&5.7& 2.6& 2.3 &  2.0 & 2.6 & 5.7\\
T$_{\rm eff-peak}$($10^5$K)&9.1&9.4&8.3&8.6&6.5&6.6 & 9.4 \\
M$_{\rm ej}$(10$^{-5}$M$_{\odot}$)&5.0& 4.8&1.8& 1.5 & .7 & 1.5& 3.3 \\
V$_{\rm max}$(km s$^{-1}$)&3563 &3681&3081& 2860 & 2772 & 3143 & 3761 \\

\enddata
\tablenotetext{a}{The initial model for all evolutionary sequences
had M$_{\rm WD}$=1.25M$_\odot$, L$_{\rm WD}$=$3.2 \times
10^{-3}$L$_\odot$, T$_{\rm eff}$=$1.9 \times 10^4$K, R$_{\rm
WD}$=3497 km, and a central temperature of $1.2 \times 10^7$K}

\tablenotetext{b} {Politano et al. (1995) library: pep reaction
{\it not} included (Weiss and Truran [1990] network); Anders and
Grevesse (1989) Solar abundances}

\tablenotetext{c} {Politano et al. (1995) library: pep reaction
{\it not} included (Hix \& Thielemann [1999] network); Anders and
Grevesse (1989) Solar abundances}

\tablenotetext{d} {Politano et al. (1995) library: pep reaction
included (Hix \& Thielemann [1999] network); Anders and Grevesse
(1989) Solar abundances}

\tablenotetext{e}{Starrfield et al. (1998) library: pep reaction
included (Hix \& Thielemann [1999] network); Anders and Grevesse
(1989) Solar abundances}

\tablenotetext{f}{Iliadis et al. (2001) library: pep reaction
included (Hix \& Thielemann [1999] network); Anders and Grevesse
(1989) Solar abundances}

\tablenotetext{g}{Iliadis 2005 library (this work): pep reaction
included (Hix \& Thielemann [1999] network); Anders and Grevesse
(1989) Solar abundances}

\tablenotetext{h}{Iliadis 2005 library (this work): pep reaction
{\it not} included (Hix \& Thielemann [1999] network); Anders and
Grevesse (1989) Solar abundances}

\end{deluxetable}

\begin{deluxetable}{@{}lccccccc}
\tablecaption{Initial Parameters and Evolutionary Results for
1.35M$_\odot$ White Dwarfs \tablenotemark{a}\label{1p35evol}}
\tablewidth{0pt} \tablehead{ \colhead{Reaction Library:} &
\colhead{P1995A} & \colhead{P1995B} & \colhead{P1995C} &
\colhead{S1998} & \colhead{I2001} & \colhead{I2005A} &
\colhead{I2005B}}\startdata

$\tau_{\rm acc}$($10^5$ yr)&2.5&3.6&2.1&2.1&2.1&1.8&3.8 \\
M$_{\rm acc}$(10$^{-5}$M$_{\odot}$)&3.9& 5.8&3.3& 3.3 &  3.3 & 2.8 &6.1\\
T$_{\rm peak}$($10^6$K)&459& 524& 413  & 414 &  407& 392& 519 \\
$\epsilon_{\rm nuc-peak}$(10$^{17}$erg gm$^{-1}$s$^{-1}$)&22.8&48.6 & 8.4& 8.6  & 4.9  & 4.4 &21.8\\
L$_{\rm peak}$ (10$^5$L$_{\odot}$)&8.0& 13.4& 9.6& 8.0 &  7.3 & 5.9 &10.9 \\
T$_{\rm eff-peak}$($10^5$K)&20.0&21.4&13&13&8.8&8.8 &18.1 \\
M$_{\rm ej}$(10$^{-5}$M$_{\odot}$)&3.3& 4.1&2.3& 2.3 &  2.3 & 1.7 &4.3\\
V$_{\rm max}$(km s$^{-1}$)&6050&7452 &5239& 4755 & 4787 & 4513 &6599 \\
\enddata

\tablenotetext{a}{The initial model for all evolutionary sequences
had M$_{\rm WD}$=1.35M$_\odot$, L$_{\rm WD}$=$4.2 \times
10^{-3}$L$_\odot$, T$_{\rm eff}$=$2.5 \times 10^4$K, R$_{\rm
WD}$=2495 km, and central temperature of $1.2 \times 10^7$K}

\end{deluxetable}

\begin{deluxetable}{@{}lccccc}
\tablecaption{Comparison of the Ejecta Abundances for
1.25M$_\odot$ White Dwarfs\tablenotemark{a} \label{1p25abund}}
\tablewidth{0pt} \tablehead{ \colhead{Sequence:} &
\colhead{P1995C} & \colhead{S1998} & \colhead{I2001} &
\colhead{I2005A} & \colhead{Init.Abund.\tablenotemark{b}}}

\startdata
H &0.30 &0.30 &0.31&0.31& 0.365\\
$^3$He$(\times 10^{-10})$&$6.8 $&$6.2$& $9.8$&$6.8$&$58000$ \\
$^4$He&0.18&0.17&0.16& 0.16 & 0.133\\
$^7$Li$(\times 10^{-8})$&$6.5$&$.94$& $22.$&$14.$& 0.0 \\
$^7$Be$(\times 10^{-8})$&$2.1$&$7.6$& $1.2$&$0.5$ & 0.0 \\
$^{12}$C$(\times 10^{-3})$  &$8.0 $&$8.6 $& $2.5$ &$4.4 $&$0.94 $ \\
$^{13}$C$(\times 10^{-3})$  &$5.6 $&$7.7 $& $1.6 $&$2.6 $& $0.012 $\\
$^{14}$N$(\times 10^{-3})$  &$5.4 $&$7.4 $& $4.8 $&$9.6 $ & $0.0023 $\\
$^{15}$N$(\times 10^{-2})$  &$7.5 $&$7.5 $& $4.1 $&$4.6 $ & $9.1 \times 10^{-5}$\\
$^{16}$O$(\times 10^{-3})$  &$13. $&$12. $& $8.5 $&$9.4 $ & 150 \\
$^{17}$O$(\times 10^{-2})$  &$3.3 $&$3.0 $& $9.4 $&$7.7 $ & $8.5 \times 10^{-5}$\\
$^{18}$O$(\times 10^{-3})$  &$3.3 $&$1.9 $& $0.4 $&$0.2 $ & $0.0048 $\\
$^{18}$F$(\times 10^{-4})$  &$6.9 $&$10. $& $1.5 $&$0.7 $ & 0.0 \\
$^{19}$F$(\times 10^{-5})$  &$5.9 $&$4.8 $& $0.8 $&$0.3 $ & $0.011$\\
$^{20}$Ne  &0.21&0.21& 0.20&0.21& 0.25 \\
$^{21}$Ne$(\times 10^{-5})$  &$8.4 $&$8.4 $& $6.4 $&$10. $& $0.09$\\
$^{22}$Ne$(\times 10^{-6})$  &$1.1 $&$0.9 $& $1.0 $&$0.2 $&$28. $ \\
$^{22}$Na$(\times 10^{-3})$ &$4.8 $&$7.3 $& $6.8 $&$4.5 $& 0.0 \\
$^{23}$Na$(\times 10^{-2})$ &$2.1 $&$1.5 $& $2.3 $&$1.9 $ & $9.2 \times 10^{-4}$\\
$^{24}$Mg$(\times 10^{-4})$ &$9.0 $&$2.5 $& $3.2 $&$2.8 $ & 1000 \\
$^{25}$Mg$(\times 10^{-2})$ &$3.3 $&$0.9 $& $1.2 $&$1.2 $& $1.9 \times 10^{-3}$\\
$^{26}$Mg$(\times 10^{-4})$ &$43. $&$19. $& $6.1 $&$6.6 $&$0.22 $ \\
$^{26}$Al$(\times 10^{-3})$ &$5.2 $&$1.9 $& $1.7 $&$2.1 $& 0.0 \\
$^{27}$Al$(\times 10^{-2})$ &$1.9 $&$1.9 $& $1.0 $&$1.0 $ & $0.0016$\\
$^{28}$Si$(\times 10^{-2})$ &$3.6 $&$5.9 $& $5.0 $&$5.5 $ & $0.018 $\\
$^{29}$Si$(\times 10^{-3})$ &$4.8 $&$5.4 $& $12. $&$10. $ & $9.5 \times 10^{-3}$\\
$^{30}$Si$(\times 10^{-2})$ &$1.6 $&$2.2 $& $2.5 $&$3.3 $ & $6.5 \times 10^{-4}$\\
$^{31}$P$(\times 10^{-2})$  &$1.0 $&$1.4 $& $1.9 $&$1.6 $ & $2.3 \times 10^{-4}$\\
$^{32}$S$(\times 10^{-2})$  &$0.6 $&$1.1 $& $1.5 $&$1.3 $ & $0.011 $\\
$^{33}$S$(\times 10^{-5})$  &$43. $&$4.4 $& $8.0 $&$7.0 $ & $0.09 $\\
$^{34}$S$(\times 10^{-5})$  &$8.3 $&$1.4 $& $2.8 $&$1.2 $ & $0.5 $\\
$^{35}$Cl$(\times 10^{-5})$  &$2.3 $&$0.7$& $1.4 $&$0.6 $ & $0.098 $ \\
$^{36}$Ar$(\times 10^{-5})$  &$2.6 $&$0.4$&$0.3$  &$0.2 $ & $1.9 $\\
$^{40}$Ca$(\times 10^{-5})$  &$1.7 $&$1.7 $& $1.7$&$1.7 $ & $1.7$\\

\enddata
\tablenotetext{a}{All abundances are Mass Fraction and are to be
multiplied by the number following the isotopic designation}

\tablenotetext{b}{Initial Abundances: half Solar (Anders \&
Grevesse 1989) and half ONeMg}

\end{deluxetable}

\begin{deluxetable}{@{}lcccc}
\tablecaption{Comparison of the Ejecta Abundances for
1.35M$_\odot$ White Dwarfs\tablenotemark{a} \label{1p35abund}}
\tablewidth{0pt} \tablehead{ \colhead{Sequence:} &
\colhead{P1995C} & \colhead{S1998} & \colhead{I2001} &
\colhead{I2005A}} \startdata

H &0.27 &0.27 &0.27&0.28 \\
$^3$He $(\times 10^{-10})$&$2.3 $&$2.1 $& $2.1 $&$1.8 $ \\
$^4$He&0.18&0.18&0.17& 0.17\\
$^7$Li $(\times 10^{-7})$ &$2.2 $&$1.2 $& $2.2 $&$1.9 $ \\
$^7$Be $(\times 10^{-8})$&$0.0$&$9.8 $& $4.1 $&$1.4 $ \\
$^{12}$C  $(\times 10^{-3})$&$8.0$&$12.$& $8.0 $&$6.2 $ \\
$^{13}$C  $(\times 10^{-3})$&$2.8 $&$4.0$& $2.4 $&$2.4 $ \\
$^{14}$N $(\times 10^{-3})$ &$4.3 $&$4.8 $& $4.3$&$8.4 $ \\
$^{15}$N  &$0.11$&0.11& $0.07 $&$0.06 $ \\
$^{16}$O $(\times 10^{-3})$ &$1.2 $&$1.1 $& $2.4 $&$2.4 $ \\
$^{17}$O $(\times 10^{-3})$ &$1.1 $&$1.0$& $59.$&$67. $ \\
$^{18}$O $(\times 10^{-3})$ &$7.8 $&$6.7 $& $3.0 $&$1.5 $ \\
$^{18}$F $(\times 10^{-3})$ &$2.5 $&$2.3 $& $0.9 $&$0.6 $ \\
$^{19}$F $(\times 10^{-5})$ &$10. $&$9.3$& $4.2$&$1.5 $ \\
$^{20}$Ne  &$0.08 $&$0.09 $& 0.11&0.12 \\
$^{21}$Ne $(\times 10^{-5})$ &$3.6 $&$3.5 $& $3.1 $&$6.2 $ \\
$^{22}$Ne $(\times 10^{-6})$ &$5.0$&$6.9 $& $4.4 $&$3.1 $\\
$^{22}$Na $(\times 10^{-2})$&$3.5 $&$5.1 $& $3.0 $&$2.3 $ \\
$^{23}$Na $(\times 10^{-2})$&$8.6 $&$6.6 $& $6.0 $&$5.8 $\\
$^{24}$Mg $(\times 10^{-3})$&$2.8 $&$1.9 $& $2.1 $&$1.9 $ \\
$^{25}$Mg $(\times 10^{-2})$&$3.8 $&$2.4 $& $2.5 $&$2.5 $\\
$^{26}$Mg $(\times 10^{-2})$&$1.6 $&$1.2 $& $0.2 $&$0.2 $ \\
$^{26}$Al $(\times 10^{-3})$&$2.8 $&$2.1 $& $2.7 $&$3.0 $\\
$^{27}$Al $(\times 10^{-2})$&$2.8 $&$3.4 $& $1.4 $&$1.4 $\\
$^{28}$Si $(\times 10^{-2})$&$2.3 $&$3.5 $& $2.7 $&$2.9 $\\
$^{29}$Si $(\times 10^{-2})$&$6.3 $&$7.0$& $19. $&$18. $\\
$^{30}$Si $(\times 10^{-2})$&$2.3 $&$2.4 $& $3.1 $&$3.8 $\\
$^{31}$P $(\times 10^{-2})$ &$3.0 $&$3.0 $& $4.3 $&$3.7 $\\
$^{32}$S $(\times 10^{-2})$ &$2.1 $&$2.8 $& $3.9 $&$4.0 $\\
$^{33}$S $(\times 10^{-3})$ &$3.4 $&$1.8 $& $2.0 $&$2.2 $\\
$^{34}$S $(\times 10^{-3})$ &$1.5 $&$1.3 $& $1.3$&$0.7 $ \\
$^{35}$Cl $(\times 10^{-3})$ &$0.9 $&$1.0 $& $1.1 $&$0.5 $ \\
$^{36}$Ar $(\times 10^{-4})$ &$6.1 $&$2.2 $& $1.5 $&$0.7 $\\
$^{40}$Ca $(\times 10^{-5})$ &$2.1 $&$2.4 $& $2.4 $&$1.8 $\\

\enddata
\tablenotetext{a}{All abundances are given as mass fraction and
are to be multiplied by the number following the isotopic
designation}

\end{deluxetable}

\clearpage

%\acknowledgements

\bibliographystyle{apj}
%\bibliography{apjmnemonic,apj_add,hix,nova,rspn_process,sn,network}

\clearpage

\begin{figure}
\plotone{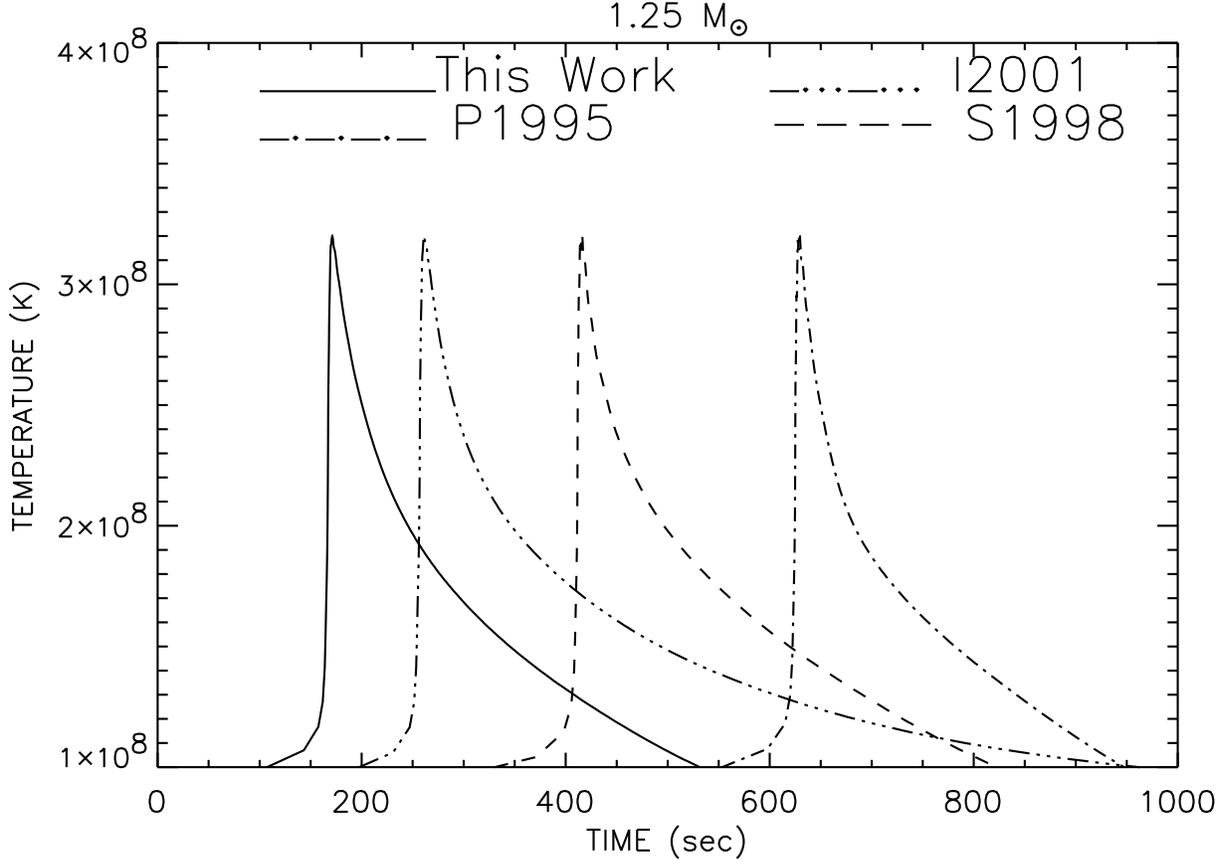} \caption{The variation with time of the
temperature in the zone in which the TNR occurs around the time of
peak temperature. For the sequences reported in this paper, this
zone is usually one zone above the core-envelope interface.  We
have plotted the results for four different simulations on a
1.25M$_\odot$ WD. The identification with calculations done with a
specific library is given on the plot. In this plot and all
following plots, S1998 refers to Starrfield et al. (1998), P1995
refers to Politano et al. (1995), I2001 refers to Iliadis et al.
(2001), and This Work refers to the calculations done with the
latest Iliadis reaction rate library (August 2005) and reported in
this paper. The details of the associated reaction rate library
are given in the text. }
\end{figure}

\begin{figure}
\plotone{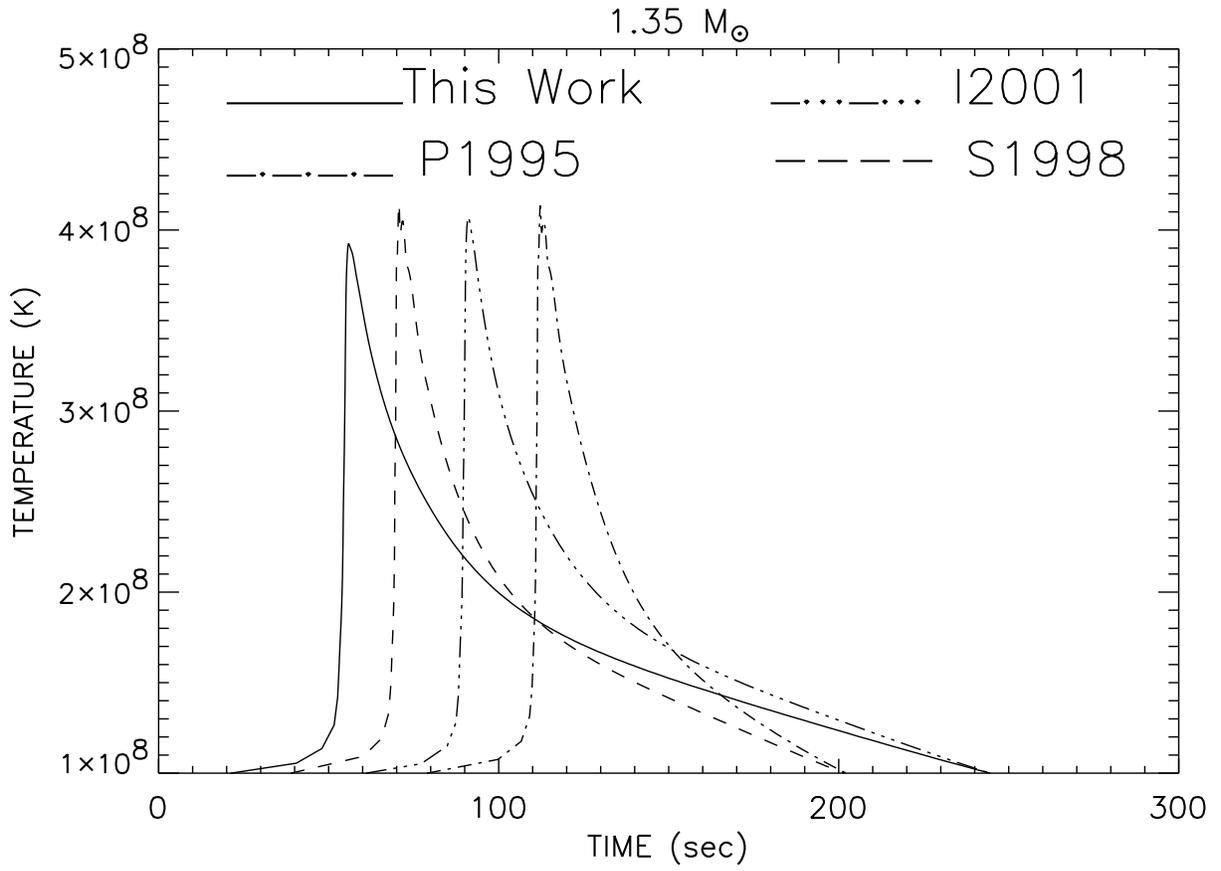}
\caption{Same as for Figure 1 but for a WD mass
of 1.35M$_\odot$.}
\end{figure}

\begin{figure}
\plotone{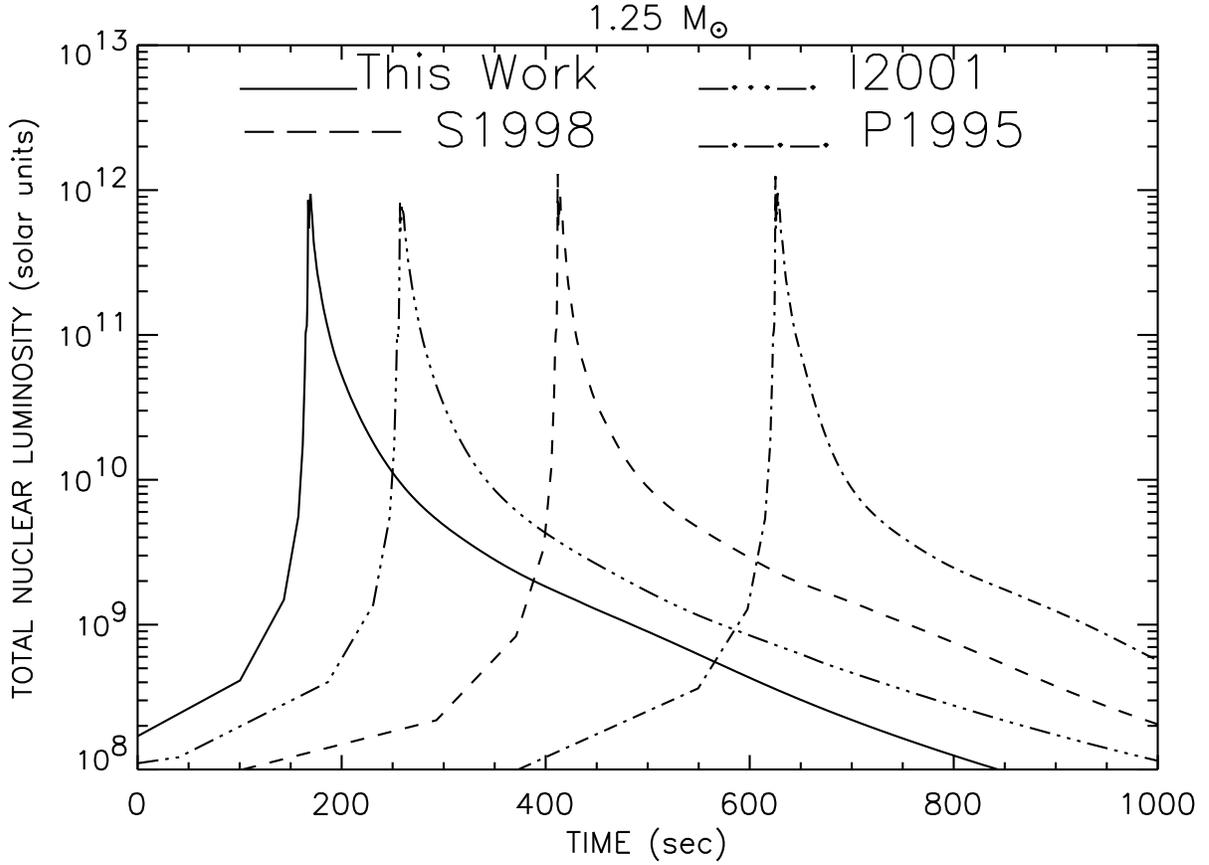} \caption{The variation with time of the total
nuclear luminosity (erg s$^{-1}$) in solar units (L$_\odot$)
around the time of peak temperature during the TNR on a
1.25M$_\odot$ WD. We integrated over all zones taking part in the
explosion. The identification with each library is given on the
plot.}
\end{figure}

\begin{figure}
\plotone{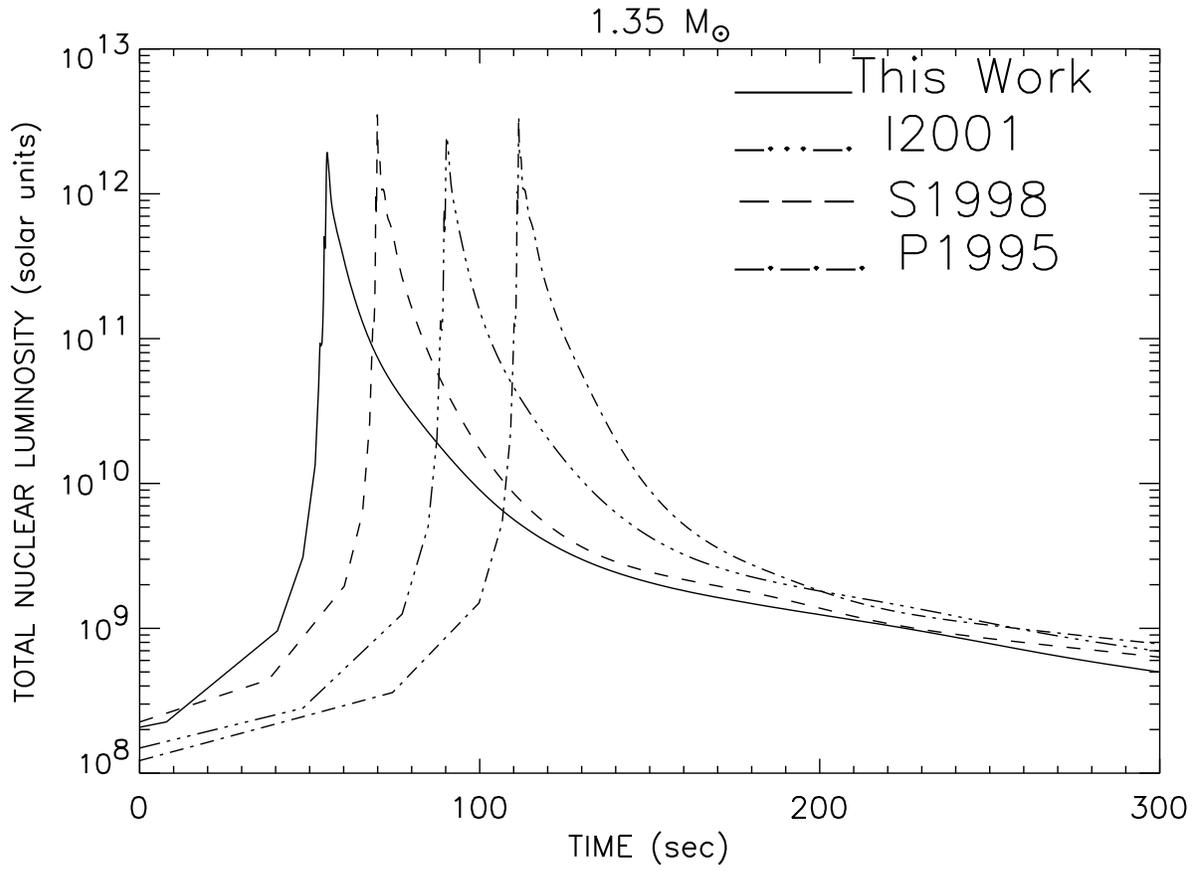}
\caption{Same as for Figure 3 but for a
1.35M$_\odot$ WD.}
\end{figure}

\begin{figure}
\plotone{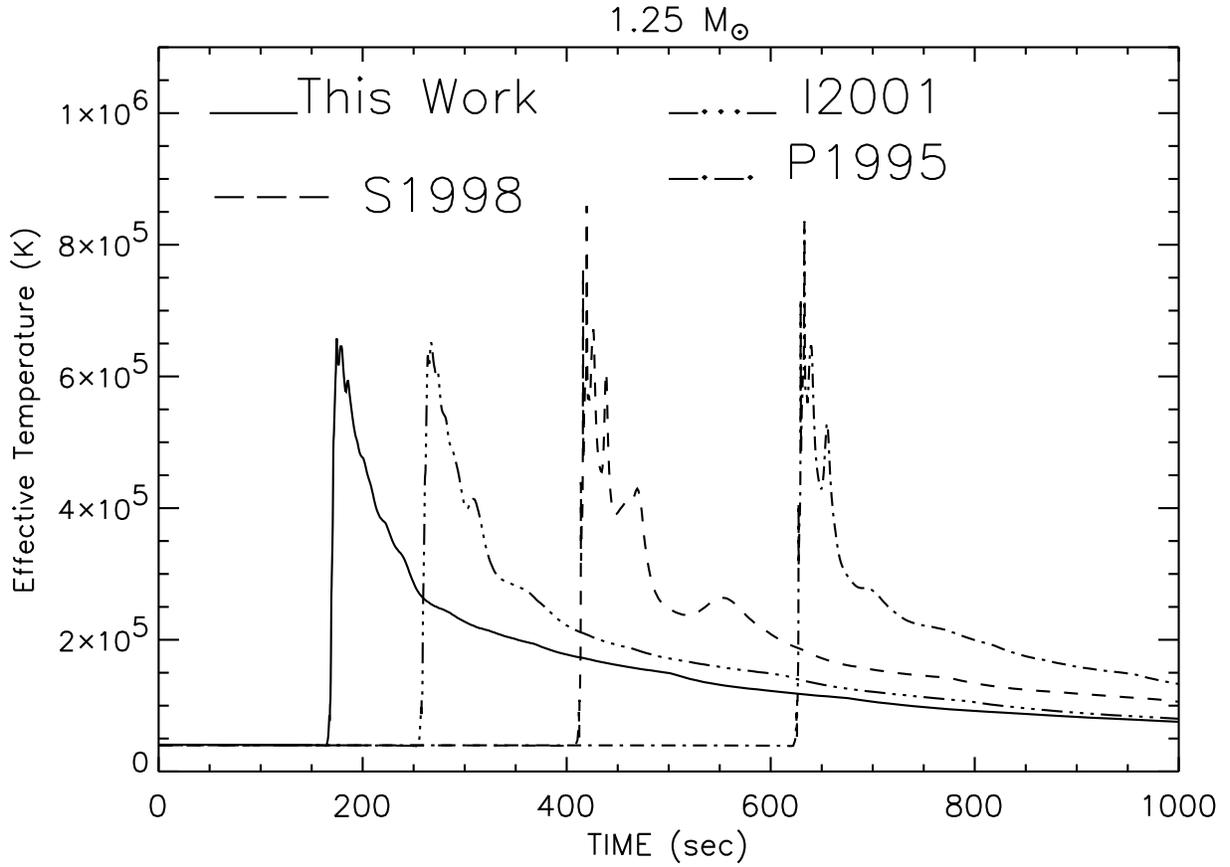}
\caption{The variation with time of the effective
temperature around the time when peak temperature is achieved in
the TNR for the sequence on the 1.25M$_\odot$ WD.  The time scale
is identical to that used in Figure 1 and shows how rapidly the
nuclear burning products are transported from the depths of the
hydrogen burning shell source to the surface.  The different
evolutionary sequences are labelled on the plot.}
\end{figure}

\begin{figure}
\plotone{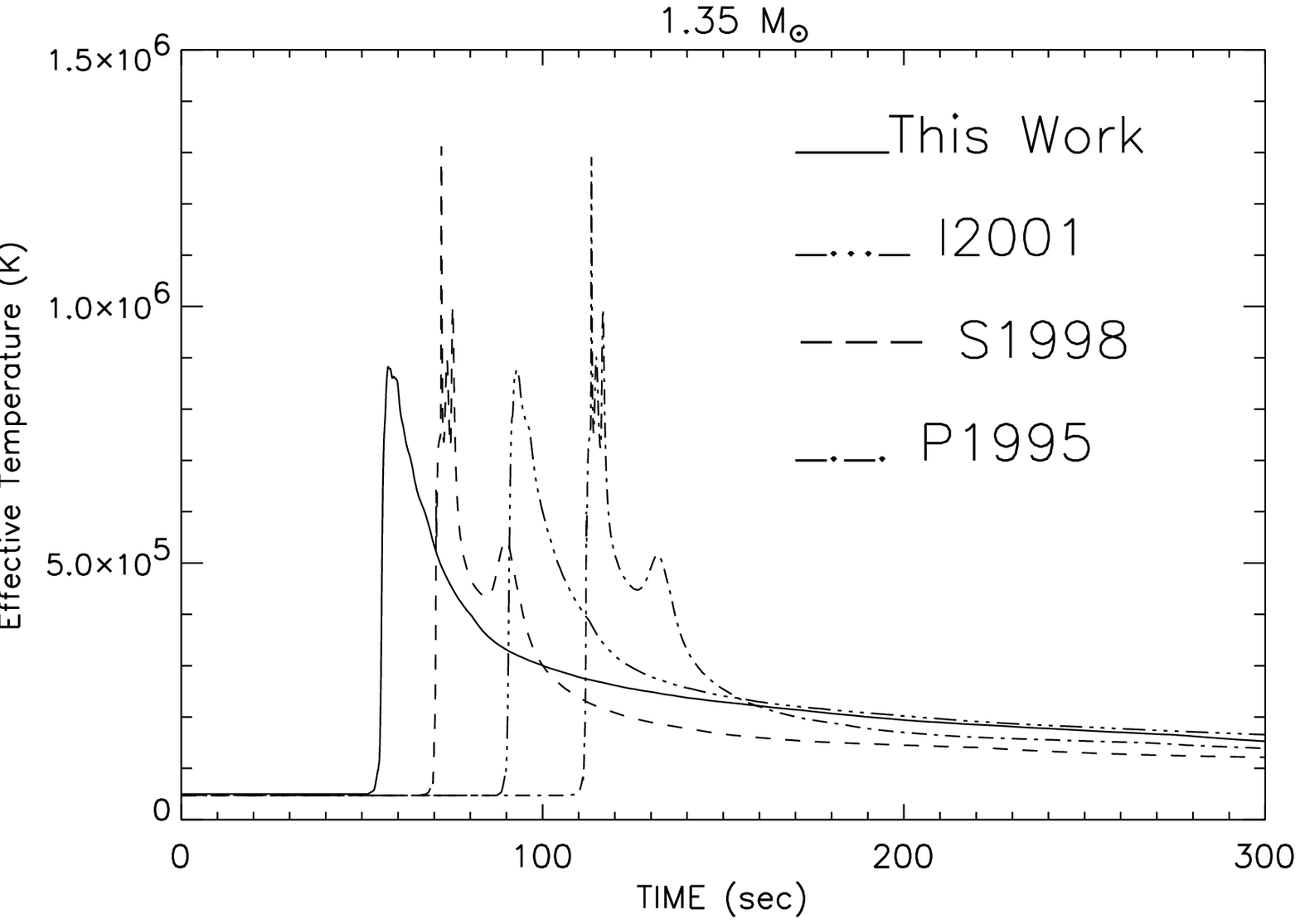}
\caption{Same as for Figure 5 but for a
1.35M$_\odot$ WD.}
\end{figure}

\begin{figure}
\plotone{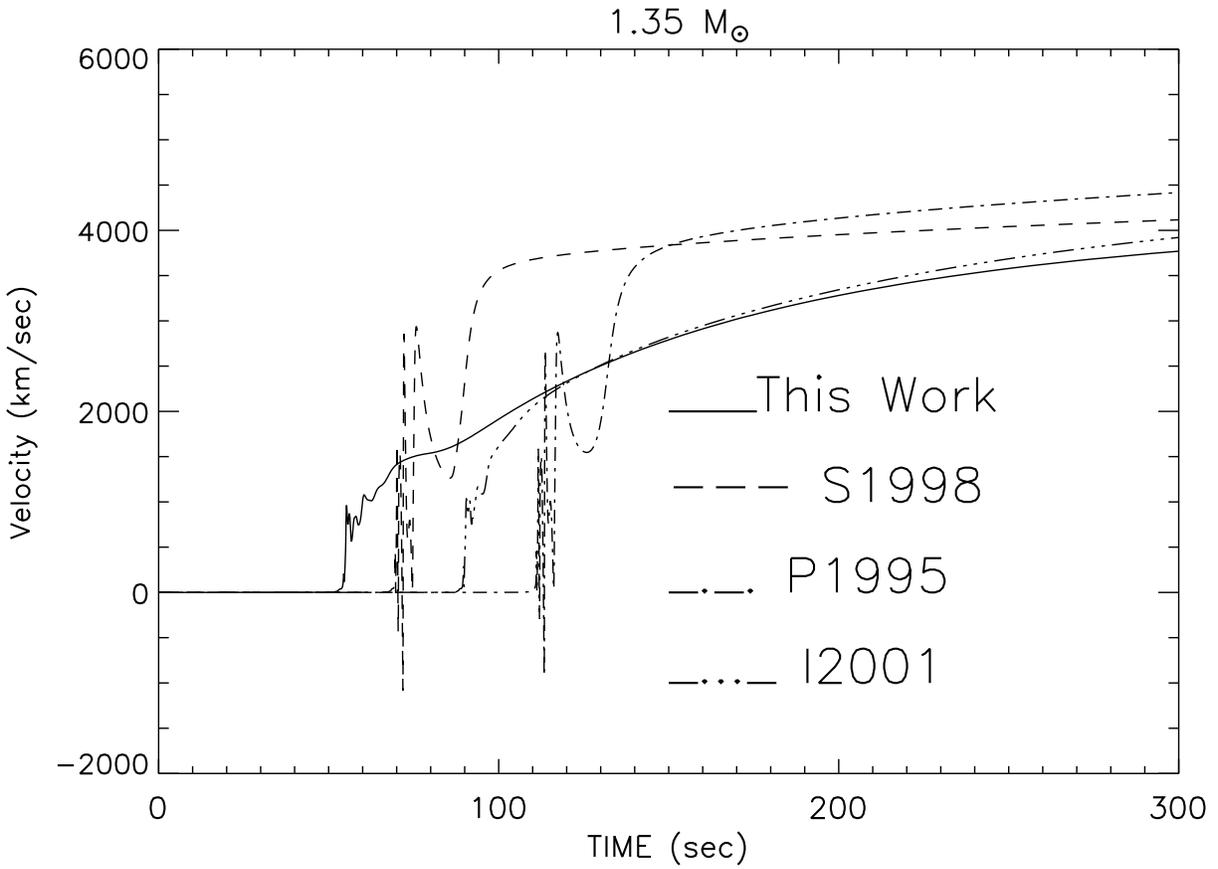} \caption{The variation with time, over the first
300 sec of the outburst, for the velocity of the surface zone
using the four different reaction libraries which are labelled on
the plot.}
\end{figure}

\begin{figure}
\plotone{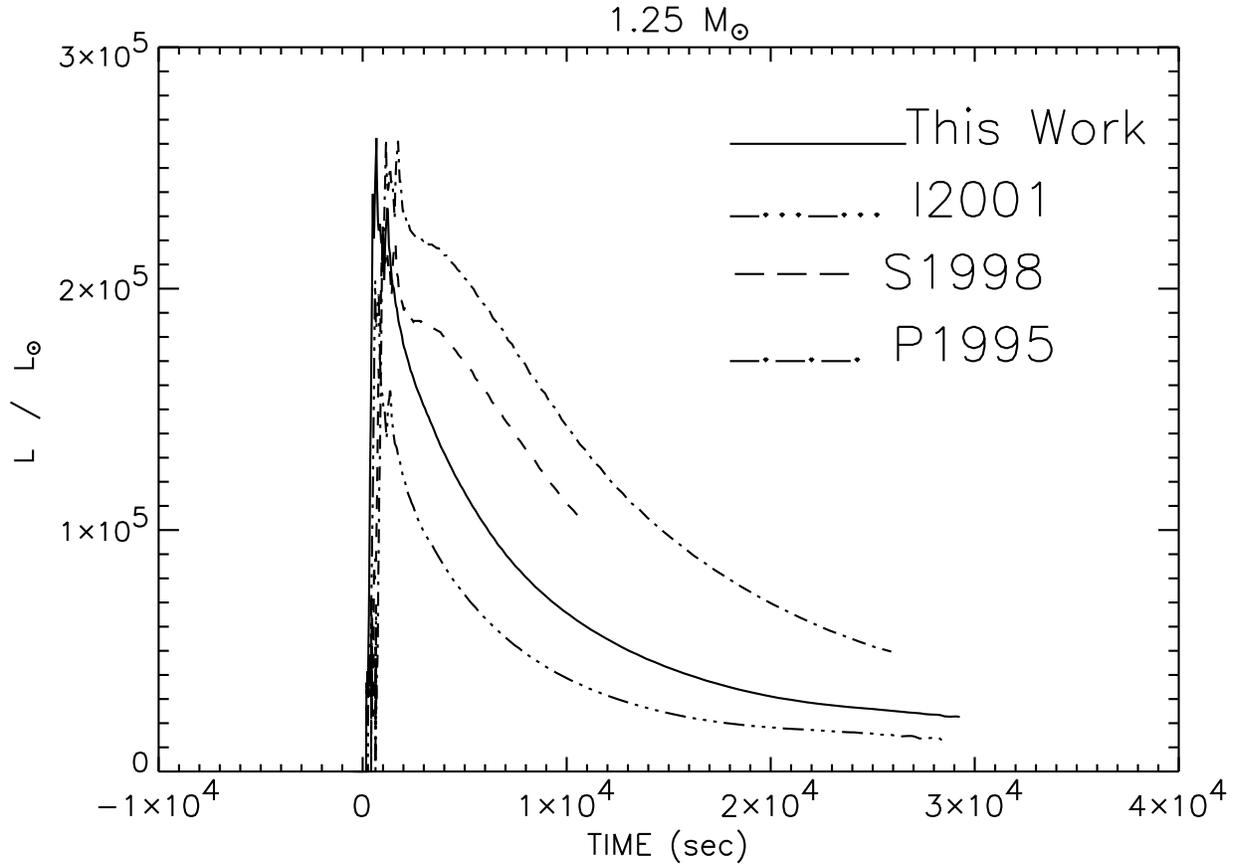}
\caption{The variation in time, over the first 11
hours of the outburst, for the surface luminosity using the four
different reaction libraries. The label which identifies each
different sequence is given on the plot. Note that as the nuclear
physics has improved, the peak luminosity and the luminosity at
later times has decreased.}
\end{figure}

\begin{figure}
\plotone{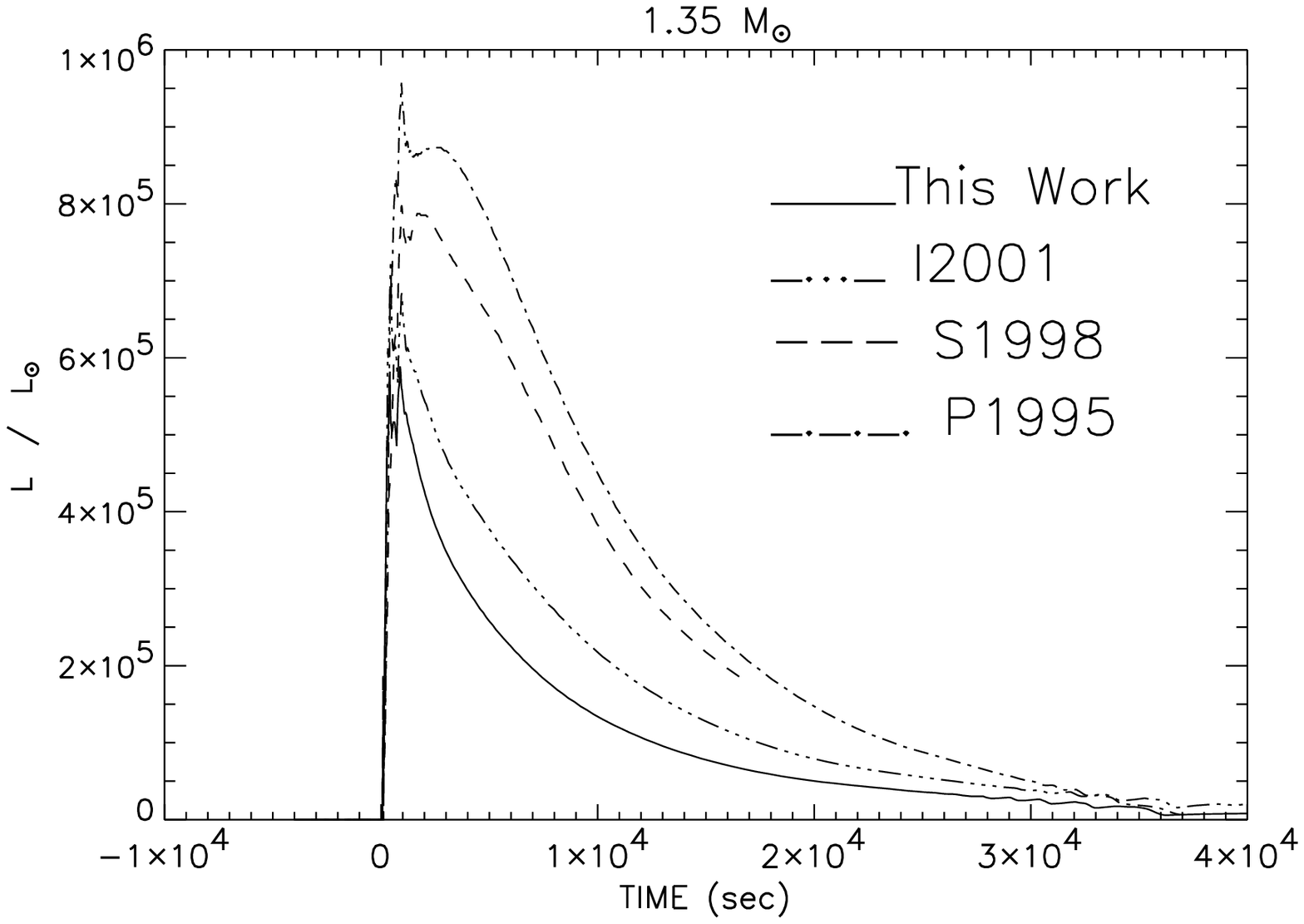}
\caption{Same as for Figure 8 but for a
1.35M$_\odot$ WD.}
\end{figure}

\begin{figure}
\plotone{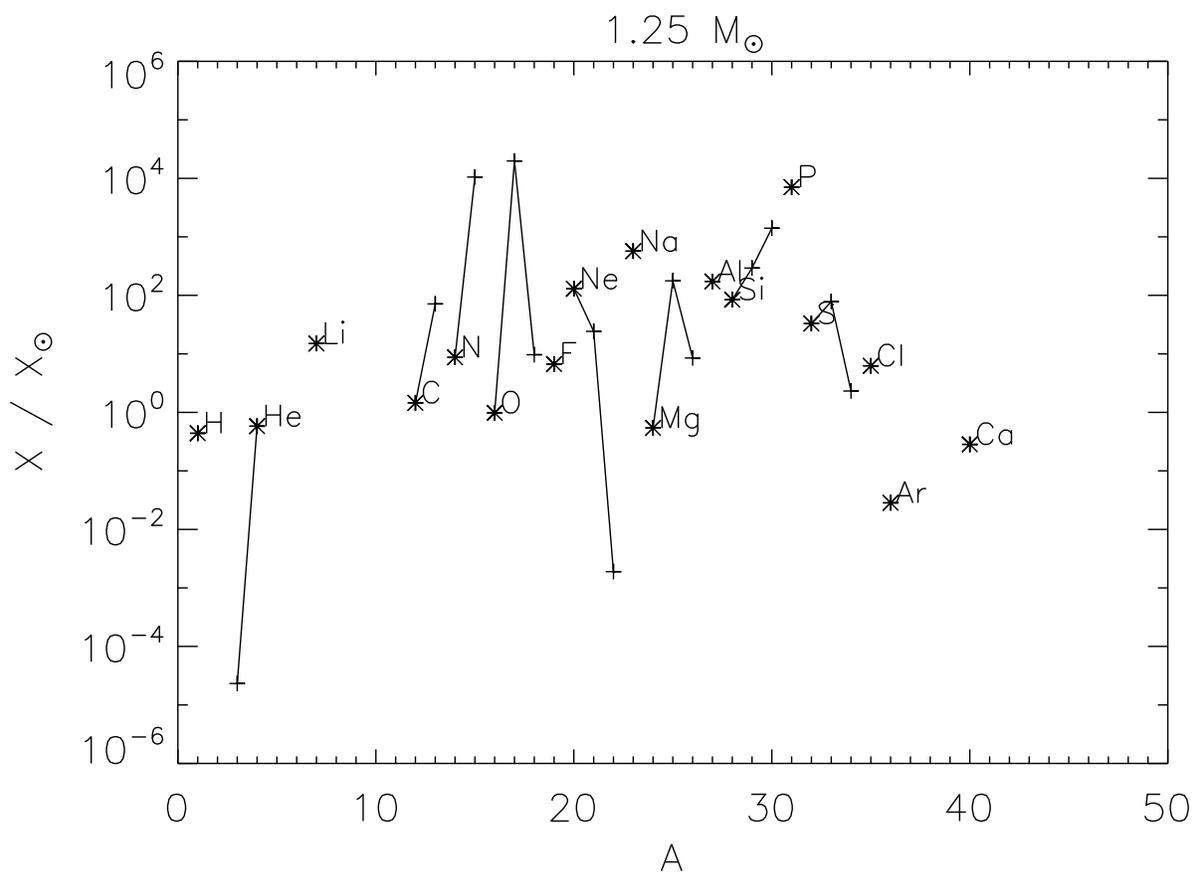} \caption{The abundances (mass fraction) of the
stable isotopes from hydrogen to calcium in the ejected material
for the 1.25M$_\odot$ sequence calculated with the I2005 reaction
rate library.  The $x$-axis is the atomic mass and the $y$-axis is
the logarithmic ratio of the abundance divided by the
corresponding Anders and Grevesse (1989) Solar abundance. As in
Timmes et al. (1995), the most abundant isotope of a given element
is designated by an ``$*$'' and all isotopes of a given element
are connected by solid lines.  Any isotope above 1.0 is
overproduced in the ejecta and a number of isotopes are
significantly enriched in the ejecta.}
\end{figure}

\begin{figure}
\plotone{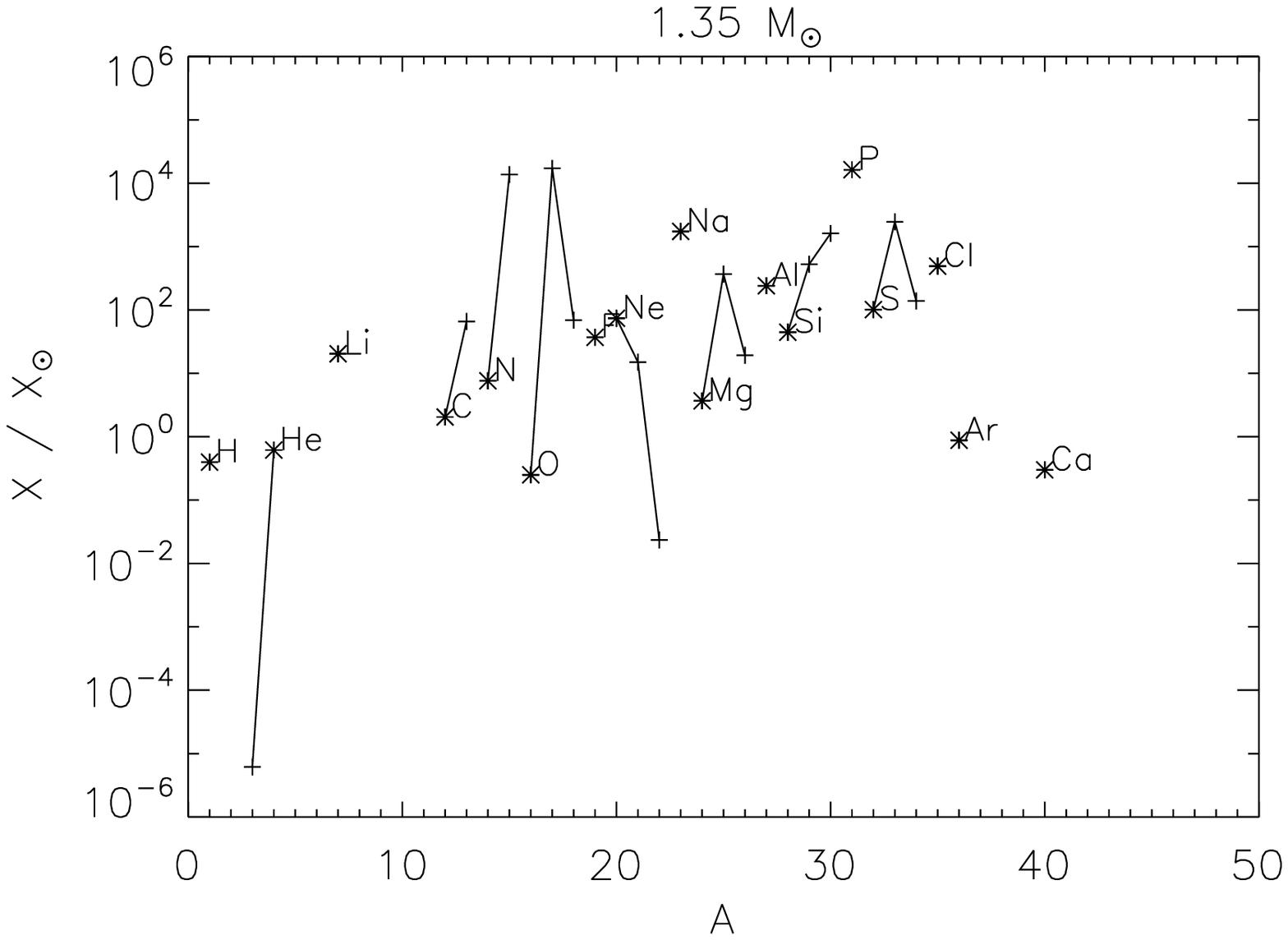}
\caption{Same as for Figure 10 but for a white
dwarf mass of 1.35M$_\odot$.}
\end{figure}

\begin{figure}
\plotone{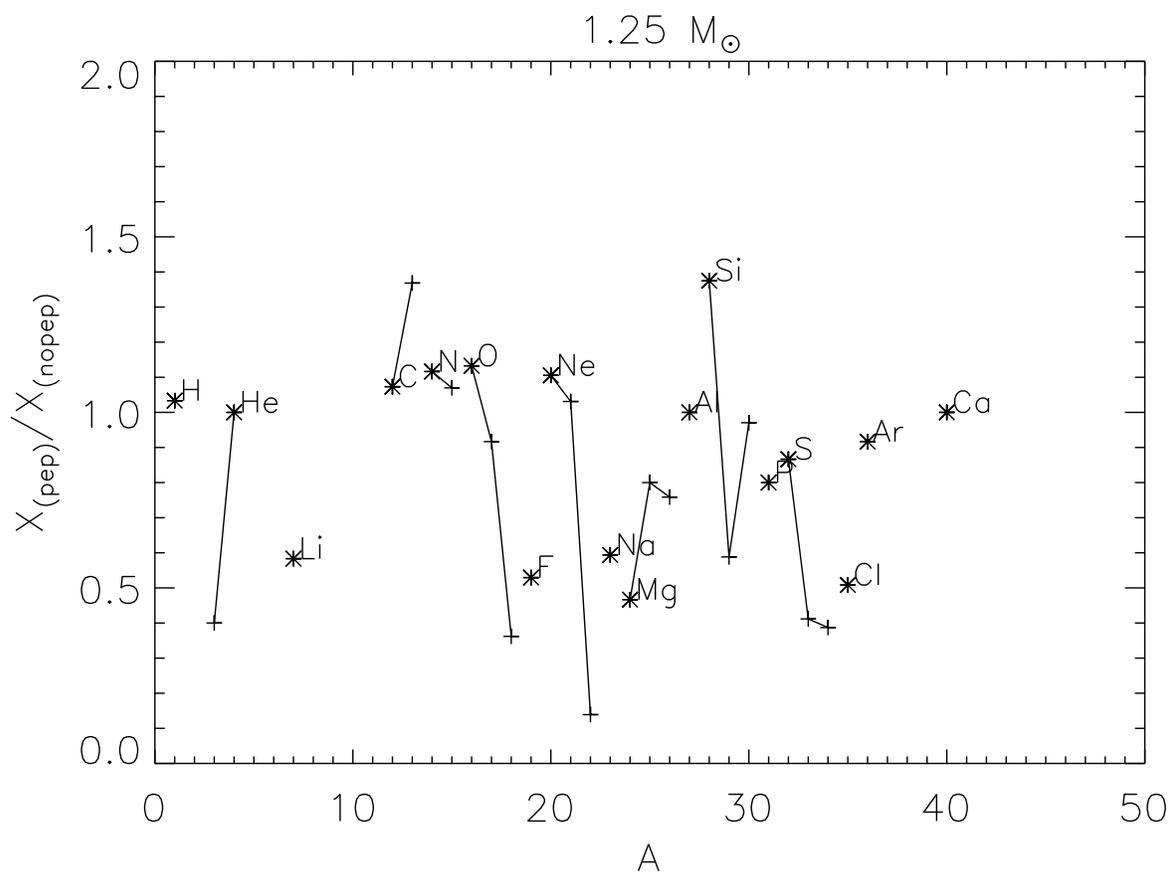} \caption{The ratio of the abundances of the
stable isotopes from hydrogen to calcium in the ejected material
for the 1.25M$_\odot$ sequences calculated with the I2005 reaction
rate library.  The $x$-axis is the atomic mass and the $y$-axis is
the linear ratio of the ejecta abundances from the sequence with
the $pep$ reaction included divided by the corresponding abundance
from the sequence calculated without the $pep$ reaction included.
The most abundant isotope of a given element is designated by an
``$*$'' and all isotopes of a given element are connected by solid
lines.}
\end{figure}

\begin{figure}
\plotone{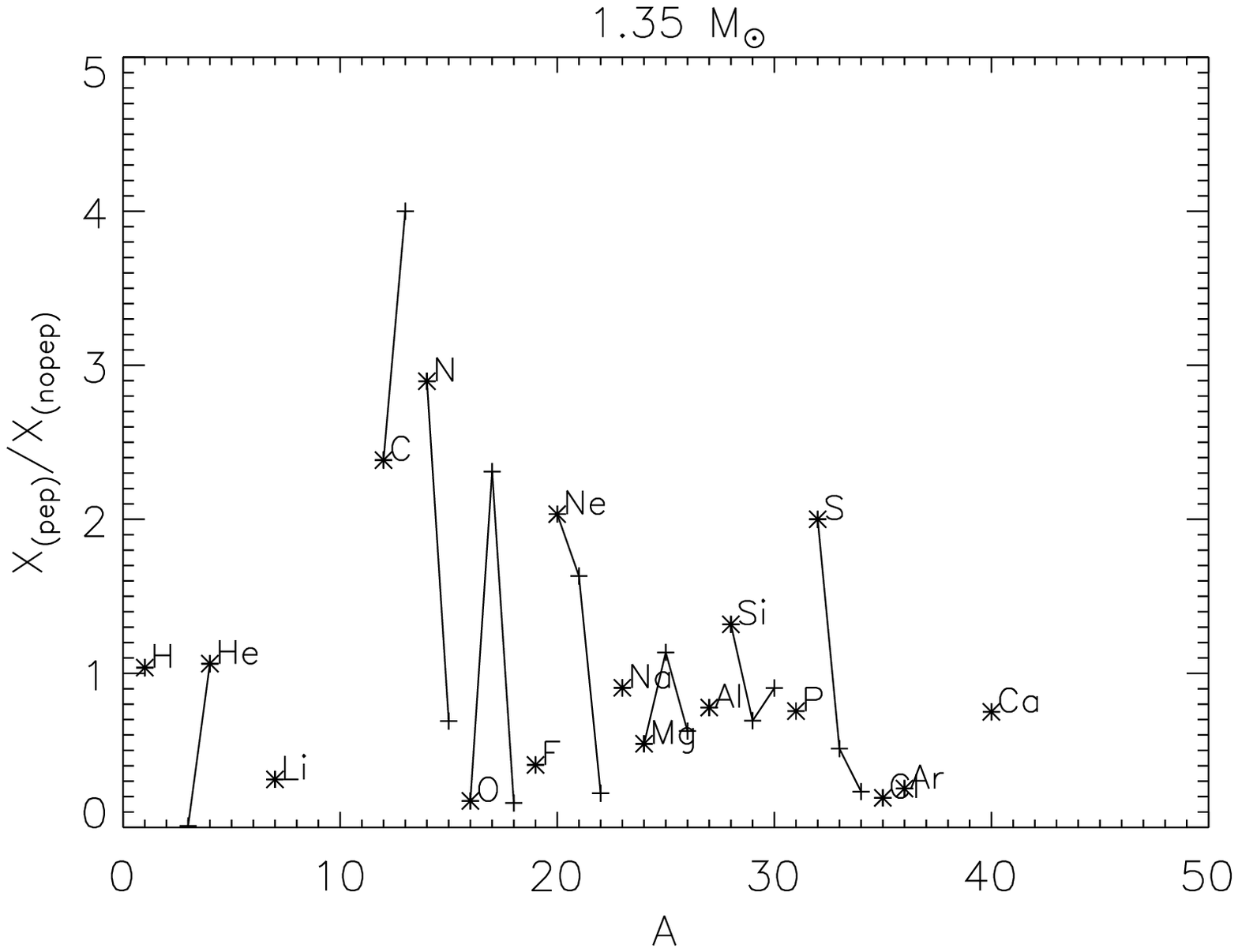} \caption{Same as for Figure 12 but for a white
dwarf mass of 1.35M$_\odot$.}
\end{figure}

\end{document}